\documentclass[11pt,rotating]{article}
\input{epsf}

\usepackage{times}
\usepackage{graphicx}

\title{Phase-field models in interfacial pattern formation out of equilibrium}
\author{
Ricard Gonz\'alez-Cinca,$^1$ Roger Folch,$^2$ Ra\'ul
Ben\'{\i}tez,$^1$
\\
Laureano Ram\'{\i}rez-Piscina,$^1$ Jaume Casademunt$^3$ and
A. Hern\'andez-Machado$^3$\\
\\
$^1$ Departament de F\'{\i}sica Aplicada,
Universitat Polit\`{e}cnica de Catalunya\\
Campus Nord, M\`odul B5, E-08034 Barcelona, Spain.\\
$^2$ Universiteit Leiden,\\
Postbus 9506, 2300 RA Leiden, The Netherlands\\
$^3$ Departament d'Estructura i Constituents de la Mat\`eria,\\
Facultat de F\'{\i}sica, Universitat de
Barcelona\\
Diagonal 647, E-08028 Barcelona, Spain}

\begin{document}
\maketitle

\section{Introduction}

A class of non-equilibrium pattern formation problems appear when an interface
moves with a velocity proportional to the gradient of some field
$u$ (which could correspond to temperature or impurity
concentration in solidification problems, pressure or another
appropriate potential in viscous fingering, etc.), which itself obeys a
bulk equation (diffusion and Laplace equations respectively in
these examples) and a Dirichlet boundary condition on the moving interface.
That constitutes a free boundary problem.
Traditional methods to numerically treat these problems imply the
explicit tracking of a sharp interface, whose dynamics is coupled
to the bulk dynamics. This can be done either by treating both
dynamics together with the corresponding moving boundary
conditions or by projecting the whole dynamics into a single
integrodifferential equation for the interface, explicitly
nonlocal and highly nonlineal.

The name {\it phase--field model} (PFM) or {\it diffuse interface
model} denotes an alternative approach to the study of such
problems. It can be seen as a mathematical tool that converts a
moving boundary problem into a set of partial differential
equations, which allows for an easier numerical
treatment. In these models an additional scalar
order parameter or phase field $\phi$ is introduced, which is
continuous in space but takes distinct constant values in each
phase. The physical interface is then located in the region where
$\phi$ changes its value, a transition layer of finite thickness
$W$. The time evolution equation  for $\phi$ is then coupled with
the $u$ field in order to take into account the boundary
conditions at the interface. When the equations are integrated the
system is treated as a whole and no distinction is made between
the interface and the bulk.
The link with the original free boundary problem is ensured by
requiring that it is recovered in the so-called sharp-interface limit
$W\rightarrow 0$.

In practice, the first phase-field models
were derived in the context of solidification
as Ginzburg--Landau time-dependent equations,
dynamically minimising a free energy functional, in the spirit of
{\it model C} of critical dynamics
in the classification of Ref. \cite{halperin:74}.  Most models for
solidification-related
phenomena continue to be derived from some thermodynamic potential.
The sharp-interface limit then serves to relate the model parameters to those
of the free boundary problem.
Nevertheless, in other phenomena (see below)
the existence of an appropiate thermodynamic
potential might not be obvious, and the PFM equations have been constructed by
directly ``guessing'' (see Sec. \ref{secpf})
what might reproduce the right free boundary problem
in the sharp-interface limit, which is then completely crucial.
From a computational point of view, such models are equally suited
to simulate the original free boundary problem as long as they do reproduce it
as $W\rightarrow 0$
\footnote{Indeed, some PFM for solidification cannot be completely
derived from a single functional ({\it non-variational} models) either,
but their discretization might converge faster (see Sec. \ref{thininterface}).}.

The earliest
formulations of a PFM from model C were done by Fix \cite{fix:83}
and Langer \cite{langer:86}. A similar PFM was introduced by
Collins and Levine \cite{collins:85}, who applied it to the study
of solidification from an undercooled melt with a diffuse
interface, showing the existence of a solvability condition for the
steady--state velocity of a flat interface for a continuous range
of undercoolings. Moreover, they provided the first link between
the PFM and the sharp--interface model with a kinetic term.
Analytical properties of the equations in Langer's PFM have been
studied by Caginalp {\it et al.}
\cite{caginalp:multi,caginalp:86a,caginalp:89,caginalp:91b},
Gurtin \cite{gurtin:87} and Fife and Gill \cite{fife:89}.
In Refs. \cite{caginalp:86a,caginalp:86-87}, Caginalp
{\it et al.} introduced anisotropic effects into a PFM. Caginalp
also showed \cite{caginalp:89,caginalp:90} that various forms of
the classical Stefan--type problem may be recovered as limiting cases of the
PFM equations when $W \rightarrow 0$. Some critical
situations where sharp--interface and PFM differ were presented in
Ref. \cite{caginalp:91a}. Penrose and Fife
\cite{penrose} provided a framework from which the
PFM equations for the case of growth under non--isothermal
conditions can be derived in a thermodynamically consistent manner
from a single entropy functional. The earlier PFM, being
obtained from a free energy functional, were only consistent
in the isothermal case.

Fix \cite{fix:83} introduced some numerical methods for the treatment of PFM
equations. After
it, a large amount of numerical computations have been carried out. Early
numerical computations
in one dimension were done by various authors \cite{1d,caginalp:91b}.
Kobayashi employed an anisotropic PFM for a pure substance in two
\cite{kobayashi:93} and
three \cite{kobayashi:94} dimensions and simulated the evolution of
dendritic--like structures growing into
an undercooled melt. The results obtained showed the computational power of this
new approach.

Wheeler {\it et al.} developed and tested thermodynamically--consistent PFM for
isothermal phase transitions in binary
alloys \cite{wheeler:92,wheeler:93b} and for the non--isothermal
case in pure substances \cite{wang:93,mcfadden:93}. Both models were used
to analyze the linear stability of a planar solidification
front in Ref. \cite{braun:94}, where a good agreement with the
free boundary problem result was found
when $W$ was taken small enough. Similar results had been obtained by
Kupferman {\it et
al.} \cite{kupferman:92} with a slightly different PFM. By combining aspects of
the models in Ref. \cite{wheeler:92}
and in Ref. \cite{wang:93}, Warren and Boettinger \cite{warren:95} constructed a
PFM for the
study of a non--isothermal binary alloy.
The behaviour of the model derived in Ref. \cite{wang:93} was
evaluated for varying $W$
and spatial and temporal resolution \cite{wheeler:93a}. This model
was also used to calculate the operating states
(tip velocity and radius of curvature) of dendrites growing from undercooled melts
and
to study the influence of anisotropic surface tension and
kinetic coefficient \cite{wang:96}. Results obtained from this model were compared
with solidification experiments
in Ref. \cite{murray:95}, and this PFM was also used to study dendritic growth in
a channel \cite{marinozzi:96}.
Moreover, it was applied to the study of mesophase growth in liquid crystals
\cite{gonzalezCinca:96,katona:96,shalaby:th,rgc:th} and it also
permitted the study of
growth with
anisotropic heat diffusion
\cite{rgc:th,gonzalezCinca:98a,borzsonyi:98}.

Kupferman {\it et al.} \cite{kupferman:94} used a PFM to study
dendritic growth in the large undercooling limit while a similar
model was used by Mozos and Guo \cite{mozos:95} in the limit of
small undercooling. More recently, some properties of the
solidification front in a supercooled liquid were derived using a
PFM \cite{bates:97}.

Fluctuations have been long introduced in the PFM in an {\it ad
hoc} way, in order to destabilize morphologically unstable
interfaces and obtain growing patterns with realistic
characteristics, for example dendrites featuring sidebranching
\cite{kobayashi:93}. Noise in PFM can also account for external
sources of fluctuations \cite{rgc:th,ricard}. In a more rigorous
approach, thermodynamical internal noise can be introduced in the
PFM to account for equilibrium fluctuations \cite{karma:99,
pavlikfluc}. This approach allows the quantitative study of
phenomena induced by internal fluctuations, like sidebranching in
solidification of pure substances \cite{karma:99}.

 Since PFM results in principle depend on the ratio $W/l_c$ of the interface
thickness to the smallest characteristic length scale of the
original free boundary problem $l_c$ as found in
\cite{wheeler:93a}, simulations had to use very small values of it
if agreement with the free boundary problem (where $W\equiv 0$)
within a few percent was desired. This became computationally
prohibitive for experimentally relevant physical parameters where
the system size is in turn much larger than $l_c$, since the
spatial discretization must resolve $W$. Although physical
interfaces are diffuse at the atomic scale, their thickness $W$
cannot attain realistical small values even in simulations of
mesoscopic structures, so that $W/l_c$ is necessarily exaggerated,
and its effects cannot be disregarded. This large range of length
scales to deal with has been the main drawback of the phase-field
method. There have been two complementary ways of circumventing
it: On the one hand, the explicit corrections to the original free
boundary problem have been computed in a few cases, either as
higher order terms in the sharp interface limit
\cite{almgren,benamar,vf1,vf2,vesicles} or in the context of the
so called thin interface limit
\cite{karma:96,karma:98,short1sided,long1sided,quanteutectic,kosterlitz}.
Some of these corrections can be cancelled out
\cite{quanteutectic,vf1,vf2,vesicles}. A complete cancellation has
so far only been achieved for the solidification of a pure
substance \cite{karma:96,karma:98} or binary alloy into a single
solid phase \cite{short1sided,long1sided}. It implies that the
simulation outcome becomes roughly independent of $W/l_c$ at some
finite value of it, and directly coincides with the original free
boundary problem within a few percent. Quantitative simulations
can then be run at that value, with no need to further decrease
$W/l_c$, which results in an enormous efficiency gain. Moreover,
the possibility of such complete cancellation has the appeal of
dealing with qualitatively distinct features such as zero kinetic
term \cite{karma:96} or zero transport coefficient in one phase
(one sided model) \cite{short1sided}.

On the other hand, large system sizes can be treated exploiting
the fact that the variation of the physical field $u$ decreases
with increasing distance to the interface. The resolution can
hence be accordingly decreased, by either using adaptive meshes
\cite{provatas:98}, or random walkers (whose amount naturally
decreases) to simulate diffusion \cite{rw}. This has enabled
quantitative comparison between theory and experiments of
dendritic solidification of a pure substance in three dimensions
\cite{tip}, showing the power of the method. To fine tune the
different physical anisotropies entering the problem, it is also
necessary to control and eventually diminish the effects of the
underlying grid anisotropy \cite{kupferman:94,karma:98,bosch:95}.

In an effort to further approach terrestrial solidification
experiments in massive samples, forced-fluid flow in two
\cite{forcedflow} and three \cite{forcedflow3d} dimensions and
natural, buoyancy-driven convection \cite{convection} have only
recently been incorporated in the phase-field formalism, usually
requiring the adaptive mesh techniques mentioned above to solve
the Navier-Stokes equations in the melt. PFM have also be
formulated for two-phase flows \cite{jacqmin}.

Although the first phase-field models were intended for the solidification
of a pure substance or binary alloy into a single solid phase, industrial alloys
usually involve more components and / or solid phases. The first models for
solidification
into more than one solid phase focussed on eutectic growth, where
two solids
of different compositions can grow from a same melt, giving rise to composite
structures which are the most common solidification patterns after dendrites.
Such first models used the standard solid--liquid phase-field and
the concentration field \cite{elder:94} or a second, solid-solid phase-field
\cite{wheeler:96} to distinguish between phases. Also using two different phase
fields, peritectic growth has more recently been studied \cite{peritectic}.
A formulation to treat an arbitrary number of phases using
as many phase fields (interpreted as volume fractions) as phases
was introduced by Steinbach {\it et al.} \cite{steinbach}.
Later on, this formulation was used in combination with a free energy
with cusp-like minima, which recovered for the first time
pure two-phase interfaces as in the
free boundary problem \cite{garcke},
removing the main difficulty to extend the progress in solidification into a
single solid phase to model multi-phase growth quantitatively
with a larger $W/l_c$ ratio.
Since such pure two-phase interfaces have recently been also achieved
for the three-phase case but using a smooth, {\em non-singular} free energy
\cite{dresden}, this possibility has become more real, and a refinement of this
model has already allowed the first quantitative simulations
in some situations \cite{quanteutectic}.

The phase-field method is also quickly expanding in different
directions: Beyond applications in solidification (for a review on
some of the topics discussed so far, see Ref.
\cite{reviewsolidif}, with a mention in solute trapping
\cite{wheeler:93b,conti:97}), PFM have been derived for other
phase transitions and transformations as those between mesophases
in liquid crystals mentioned above
\cite{gonzalezCinca:96,katona:96,shalaby:th}, solid-state
structural phase transformations, and electrodeposition. Other PFM
with an impact on materials science include nucleation
\cite{roy:98}, grain growth \cite{grains} and coarsening, domain
evolution in thin films, 
spiral growth \cite{kam}, and models incorporating
elastic effects, such as those for stress-induced instabilities
\cite{stress-induced}, the dynamics of dislocations
\cite{dislocations} and the propagation of cracks \cite{fracture}.
(For a short review on some of those, see Ref.
\cite{reviewmaterials}).

Further outstretching their reach, some PFM have recently been
derived for the dynamics of fluid interfaces. Because such cases
are examples of interfaces between media which are not necessarily
different thermodynamic phases of a same substance, the existence
of a free energy functional to minimize is not obvious. The first
model for viscous fingering including arbitrary fluid viscosities,
for instance, was derived purely on the basis of reproducing the
desired free boundary problem in the sharp-interface limit,
without the guide of any free energy functional \cite{vf1,vf2}. In
contrast, later formulations of the same problem were derived from
such a functional \cite{variationalvf}. This nicely illustrates
two crucial points: On the one hand, given a free boundary
problem, a PFM for it is not unique, which allows for the
necessary freedom to construct models cancelling some of the
finite interface thickness corrections
\cite{almgren,benamar,vf1,vf2,vesicles,karma:96,karma:98}
 to that
free boundary problem, yielding more accurate and computationally
efficient models, as discussed above. On the other hand, a PFM can
derive or not from a free energy functional. We emphasize this
latter point, because we believe that removing this restriction
might help derive models for other problems. An even more recent
example of that is a PFM for vesicle dynamics \cite{vesicles},
based on the viscous-fingering model of Refs. \cite{vf1,vf2} and
therefore not derived from a free energy functional.

Finally, let us mention that the key step of the PFM, the
calculation which relates it to a specific free boundary problem,
can sometimes also be applied to phenomena with a physical diffuse
interface
in order to obtain an effective sharp-interface description of them for
theoretical
purposes. One example of those are the steady states of
thermal plumes, where finite
interface thickness corrections to the resulting free boundary problem
explained the selection mechanism for their width \cite{benamar}.
Not all diffuse-interface problems can be mapped to an effective free boundary
problem. It has been shown that the so-called pulled fronts cannot
\cite{pulled}, and other
singular fronts require a generalisation of the method employed \cite{bacteria},
as
those arising in reaction--diffusion systems with a non-linear diffusivity,
e.g. in physical models of bacterial colony growth \cite{bacteria}.

The main features of PFM that make them adequate to describe
interfacial
dynamics can be summarized
as follows:
\begin{itemize}

\item
As sets of coupled partial differential equations, they are
simpler to integrate numerically than the integrodifferential type
of equation usual in sharp-interface descriptions. (In particular,
they do not suffer from long-time numerical instabilities as some
of the latter). This also makes modifications of the model
comparatively easier to incorporate, e.g. to include anisotropy
\cite{kupferman:94,karma:98,bosch:95}, external fluctuations
\cite{kobayashi:93,ricard} or fluid flow
\cite{forcedflow,forcedflow3d,convection,jacqmin}.

\item PFM are especially suited for the introduction of internal
fluctuations, since these are usually known from first principles
\cite{karma:99}.

\item More specifically, PFM have the well-recognised advantadge
that the location of the interface does not have to be explicitly
determined as part of the solution, but it is obtained {\em from}
the solution of the equations for the phase field(s). Not having
to track the interfaces, topological changes of the interface
(self--intersections, coalescence or pinching of droplets, etc.)
and extensions to three dimensions are more feasible within the
PFM formalism. (Sharp-interface formulations are normally
restricted to connected interfaces and to two dimensions, either by
the formulation itself or because they become typically impractical 
in those other cases).

\end{itemize}

Recently, the concept of PFM has been used in a broader sense to
include any model which contains continuous fields that are
introduced to describe the phases present with diffuse interfaces
\cite{kosterlitz, mexico, castro}. In this context models for
order-disorder transitions, spinodal decomposition and
heterogeneous nucleation \cite{kosterlitz, castro}, described by
only one equation for a scalar field,  has been included at the
same level than dendritic growth and solidification of eutectic
alloys. In Ref. \cite{mexico} a continuous one-sided model has
been proposed to model Hele-Shaw flows in the high viscosity
contrast regime. In the same spirit, continuous models have been
introduced in the field of roughening due to their nonlocal
properties \cite{ala, euro}.

In this chapter we review some results concerning the use of PFM
in the context of pattern formation in morphologically unstable
interfaces. More specifically, we address two problems which have
played a prototypical role in this context, namely solidification
and viscous fingering. Whereas PFM for solidification are usually
derived from a free energy functional, this is not necessarily the
case for viscous fingering, as mentioned above. This will enable
us to illustrate the two situations: In Sec. \ref{solidification}
we derive a PFM for solidification from a free energy functional;
in Sec. \ref{viscousfingering}, we discuss the appropriate PFM for
viscous fingering which directly gives the sharp interface model.
In both cases the asymptotic calculation that relates the PFM to
the sharp interface counterpart is a central step to validate the
model. In the next sections we will show as examples results of
applying PFM to different situations: fluctuations in
solidification, growth of liquid crystal mesophases which are
characterized by strong anisotropies, and viscous fingering in
Hele-Shaw cells.

\section{A first example: solidification of a pure substance}
\label{solidification}

\subsection{The Sharp Interface Model}
\label{stefanproblem}

The standard sharp--interface macroscopic model for free solidification of a pure
substance \cite{langerrmp} considers the solid--liquid interface as a microscopically thin moving
surface.
This model relies on the heat diffusion equation together with two boundary
conditions at the interface, which is
assumed to be sharp. The diffusion equation for the temperature field $T$ is given
by
\begin{equation} \label{diff}
\frac{\partial T}{\partial \tilde{t}} = D \tilde{\nabla}^2 T,
\end{equation}
where $\tilde{t}$ is time and $D$ is the
diffusion coefficient ($D=k/c_{p}$, being $k$ the
heat conductivity and $c_{p}$
the specific heat per unit volume).
Tilded variables denote magnitudes in physical units.
We are considering the symmetrical model of
solidification, that is,
we assume that $D$ and $c_{p}$ are equal in both phases.

One boundary condition is obtained from the conservation of the released latent
heat at the moving interface:
\begin{equation} \label{cons}
L\tilde{\upsilon}_{n} = D c_{p} [(\tilde{\nabla} _n T)_{S}-
(\tilde{\nabla} _n T)_{L}],
\end{equation}
where $L$ is the latent heat per unit volume,
$\tilde{\upsilon}_{n}$ is the normal velocity of the interface,
$\tilde{\nabla}_n$ is the normal derivative at the interface and S
and L refer to solid and liquid respectively. The left--hand--side
corresponds to the rate at which heat is produced at the interface
per unit area while the right--hand--side is the total energy flux
away from the interface. The sharp interface model is completed
with a thermodynamic boundary condition which assumes local
equilibrium at the interface and incorporates the non-equilibrium
effect of the kinetic attachment of atoms at the moving interface:
\begin{equation} \label{gibbs0}
T_{int} =
T_{M}-\frac{T_{M}\sigma}{L}\tilde{\kappa}-\frac{\tilde{\upsilon}_{n}}{\mu},
\end{equation}
where $T_{int}$ is the temperature at the interface, $\sigma$ is
the surface tension, $\mu$ is the kinetic coefficient and
$\tilde{\kappa}$ is the local curvature of the interface, being
positive when the solid bulges into the liquid. This relation can
be thermodynamically derived by considering that at local
equilibrium the solid and liquid Helmholtz free energies are
equal. The ratio $\sigma/L$ has dimensions of a length which sets
the scale of the pattern. If $\sigma=0$ the so called Stefan
problem is recovered and no realistic pattern formation can be
described. Introducing the reduced diffusion field
$u=(T-T_{M})/(L/c_{p})$ and scaling Eqs. (\ref{diff}-\ref{gibbs0})
with the diffusion length $l=D/\upsilon_{c}$ and the diffusion
time $\gamma=l^2/D$ ($\upsilon_{c}$ is some characteristic front
velocity, in most experiments in the range of $1 \mu m/s$), we
obtain a dimensionless set of sharp-interface equations
\begin{eqnarray} \label{si-dif}
&&\frac{\partial u}{\partial t} = \nabla^2 u,
\\
\label{si-cons} &&\upsilon_{n} =  [(\nabla _n u)_{S}-(\nabla _n
u)_{L}],
\\
\label{si-gt} &&u_{int} =
-\frac{d_{0}}{l}{\kappa}-\frac{\beta}{\gamma/l} \upsilon_{n},
\end{eqnarray}
where $d_{0}=T_{M}\sigma c_{p}/L^2$ is the capillary length of the
substance and $\beta = c_{p} /( \mu L)$.

\subsection{Derivation from a free energy functional}
\label{fromfunctional}

PFM consist in a set of equations describing the dynamics of a
continuous order parameter $\phi({\bf r},t)$ which takes different
constant values in the solid and liquid phases. The solidification
front is then determined by the implicit condition $\phi({\bf
r},t)=$ {\it const}. , and the transition between solid and liquid
phases takes place in a diffuse region of thickness $W$. The
dynamics of the phase-field $\phi({\bf r},t)$ is coupled with the
evolution of the diffusion field, and can be derived from a
free-energy functional such as
\begin{equation}
F[e,\phi]=\int d{\bf \tilde{r}} \left[ K|{\bf
\tilde{\nabla}}\phi|^2 + h_0f(\phi) + e_0u^2 \right],
\label{eq:free-energy}
\end{equation}
where $u=e/e_0 + h(\phi)/2$, being $e$ the local enthalpy per unit
volume, $e_0=\frac{L^2}{T_{M}c_{p}}$, and $h$ a certain polynomial
function of $\phi$. The gradient term in the functional takes into
account the free energy cost of interfacial deformations and $f(\phi)$
is a double-well potential describing the bulk free-energy
density. The minima of $f(\phi)$ determine the constant values of
$\phi$ at the bulk phases. The constants $K$ and $h_{0}$ have
dimensions of energy per unit length and energy per unit volume
respectively.

The basic phase-field equations can be then obtained by applying
the well known variational relations
\begin{eqnarray}\label{variational1}
\partial_{\tilde{t}} \phi &=& -\frac{1}{\tau}
\left(\frac{\delta F}{\delta \phi} \right) + \tilde{\eta}({\bf
\tilde{r}},\tilde{t})
\\
\partial_{\tilde{t}} e &=&
De_0\tilde{\nabla}^2\left(\frac{\delta F}{\delta e} \right) - {\bf
\tilde{\nabla}}\cdot{\bf \tilde{q}}_e({\bf \tilde{r}},\tilde{t}).
\label{variational2}
\end{eqnarray}

These are the equations of model C, applicable to the dynamics of
two fields, one conserved and the other nonconserved
\cite{halperin:74}. The terms $\tilde{\eta}({\bf
\tilde{r}},\tilde{t})$ and ${\bf \tilde{q}}_e ({\bf
\tilde{r}},\tilde{t})$ are noises accounting for thermal
fluctuations. These noise terms can be ignored if fluctuations are
not relevant in the study at hand.

After some calculations and scaling space and times
with $l,\gamma$, we get the set of equations
\begin{eqnarray} \label{pf1}
\alpha \varepsilon^2\partial_{t} \phi &=&
\varepsilon^2\nabla^2\phi- f'(\phi) - \frac{e_0}{h_0}
h'(\phi)u+\eta({\bf r},t)
\\
\partial_{t}u &=& \nabla^{2}u+\frac{1}{2}\partial_{t}h -
{\bf \nabla}\cdot {\bf q_{u}}({\bf r},t) \label{u1}
\end{eqnarray}
where  $\varepsilon=\frac{W}{l}$, $W=\sqrt{K/h_0}$ and
$\alpha=\frac{D\tau}{W^2h_0}$. The statistical properties of the
noises can be determined by considering the
fluctuation-dissipation theorem in Eqs. (\ref{variational1},
\ref{variational2})
\begin{eqnarray}\label{eta}
\langle \eta({\bf r},t)\eta({\bf r}',t') \rangle &=& \frac{2
K_bT_M\tau}{h_{0}^2 \gamma l^d} \delta({\bf r}-{\bf r'})
\delta(t-t')
\\
\langle q^i_u({\bf r},t) q^j_u({\bf r}',t') \rangle &=& \frac{2
K_B T_M D e_0}{\gamma l^{d+1}} \delta_{ij}  \delta({\bf r}-{\bf
r'}) \delta(t-t'). \label{q}
\end{eqnarray}

\subsection{The thin-interface limit}
\label{thininterface}

In this subsection we will perform the $W\rightarrow 0$ limit of a
PFM for solidification. Our aim is first to obtain the
relationship of the PFM parameters with those of the sharp
interface model, by using the thin interface asymptotics. Second
we want to illustrate how can one construct a PFM with the only
criterium of reproducing a desired sharp interface model. In
particular we will use a variant of Eqs. (\ref{pf1},\ref{u1})
derived
above, not necessarily obtainable from any energy functional.

Consider the general
model equations
\begin{eqnarray} \label{pf2}
\alpha \varepsilon^2\partial_{t} \phi &=&
\varepsilon^2\nabla^2\phi- f'(\phi)-\lambda g'(\phi)u
\\
\partial_{t}u &=& \nabla^{2}u+\frac{1}{2}\partial_{t}h
\label{u2},
\end{eqnarray}
where $\lambda=\frac{2e_0}{bh_0}$ and
no particular choice for the function $h$ has been taken.
For the particular choice of $h(\phi)=\frac{2}{b}g(\phi)$ with
$b=g(1)-g(-1)$, Eqs. (\ref{pf2}, \ref{u2}) recover the variational
structure described in the last section.
Other choices for the function $h$ improve the computational
efficiency of the model but the variational properties
of the equations are lost. The non-variational formulation of
the problem has some thermodynamical drawbacks:
For instance, the fluctuation-dissipation theorem
loses its validity and the noise properties should be calculated by
means of other procedures \cite{raul2}.
In any case, the only restriction which must be imposed to the
function $h$ is that it satisfies $h(\pm 1)=\mp 1$.
Usual choices for the free-energy density potential $f$
and for the coupling function $g$ are
\begin{eqnarray}
f(\phi) &=& \frac{\phi^4}{4}-\frac{\phi^2}{2}\\
g(\phi) &=& \phi-\frac{2\phi^3}{3}-\frac{\phi^5}{5}.
\label{eq:potentials}
\end{eqnarray}
We present the thin-interface limit without
considering the noise terms (the extension to fluctuating PFM will
be presented elsewhere \cite{raul2}).

As we mentioned in the introduction, the phase-field Eqs.
(\ref{pf1},\ref{u1}) reproduce the sharp-interface problem when
the interface thickness $W$ is small enough compared with the
smallest physical length scale, which in solidification is the
capillary length $d_0$. However, Karma and Rappel introduced the
so-called {\it thin-interface limit}, in which they assumed
$\varepsilon=\frac{W}{l}\ll 1$, but not necessarily $W<d_0$
\cite{karma:98}. This corresponds to take the limit
$\epsilon\rightarrow 0$ while keeping $\lambda$, which we will see
to control the ratio $W/d_0$, constant.  Of course, it introduces
corrections at first order in $W/d_0$ coming from the term
$-\lambda g'(\phi)u$. The point is that, for $h(\phi)$, $g(\phi)$
and $f(\phi)$ odd, these corrections can be tracked to change {\em
only} the value of $\beta$ being simulated. Once one has
identified the new $\beta$, the model parameters can be adjusted
to reproduce the desired $\beta$ while keeping $W$ even larger
than $d_0$ in practice\footnote{As a side benefit, $\beta$ can now
be made to vanish, because these corrections can have the opposite
sign that the $W/d_0\rightarrow 0$ contribution.}. This is
completely equivalent to perform a more traditional asymptotic
expansion using $W/d_0$ as a small parameter but keeping
corrections up to first order in $W/d_0$ as proposed in the
introduction. Indeed, corrections to the mass conservation
Eq. (\ref{cons}), not coming from the  $-\lambda g'(\phi)u$,
are ignored in the $W/l\rightarrow 0$ limit. They turn out to
vanish for $h(\phi)$ odd in the two-sided case considered here
where $D$ is the same in both phases, but not in the one-sided
case where the solid diffusivity is negligible compared to that of
the liquid \cite{almgren,short1sided,long1sided}. Here, we follow
the thin-interface limit in Ref. \cite{karma:98}, as the most
compact way of obtaining the correct result. For a higher order
$W/d_0$ expansion, see the appendix of Ref. \cite{karma:98},
\cite{almgren} (for unequal diffusivities) or \cite{long1sided}
(for the strict one-sided case) for more detailed presentations.
See also Ref. \cite{kosterlitz}. An expansion without any {\it a
priori} assumption on which physical scale to compare the
phase-field small parameters with can also be found in Ref.
\cite{vf1} in the context of viscous fingering.

In order to perform the asymptotic expansion, we divide our system
in two different regions: an outer region at the bulk phases far
away from the interface and an inner region located around the
interface at a distance W. We will use capital letters to refer to
all the fields in the inner region. In the outer region we expand
both fields for $\varepsilon \ll 1$
\begin{eqnarray}
u &=& u_0+\varepsilon u_1+O(\varepsilon^2)\\
\phi &=& \phi_0 + \varepsilon\phi_1+O(\varepsilon^2).
\label{outer}
\end{eqnarray}
Introducing Eq. (\ref{outer}) into Eq. (\ref{pf2}) we obtain, at order
zero in $\varepsilon$
\begin{eqnarray}
&&-f'(\phi_0)-\lambda g'(\phi_0)u_0 = 0. \label{outer0-phi}
\end{eqnarray}
It is easy to see that $\phi_0=\pm1$ is a solution of the last equation
for our particular choice of the functions $f(\phi)$ and $g(\phi)$.
As $h$ is constant in the bulk, the outer solution of the
diffusion field at zero order is given by the diffusion equation
\begin{eqnarray}
\partial_t u_0 &=&\nabla^2 u_0.
\end{eqnarray}
Higher orders in $\varepsilon$ are described by
\begin{eqnarray}
\phi_i &=& 0,
\\
\partial_t u_i &=&\nabla^2 u_i\;,  i>0.
\end{eqnarray}
In order to work in the inner region, we consider Eqs. (\ref{pf2})
and (\ref{u2}) and write them in a three dimensional orthogonal
curvilinear coordinate system $(r,s,w)$ based in the surface
defined by $\phi({\bf r},t)=0$. The normal coordinate of this
coordinate system $r$ is then scaled using the small parameter
$\varepsilon$ so that $\rho=\frac{r}{\varepsilon}$. Writing the
equations in the frame of the moving front and keeping terms until
first order in $\varepsilon$ we have
\begin{eqnarray}
\alpha v\varepsilon \partial_\rho \Phi&=&\partial^2_\rho \Phi -f'
+\varepsilon\kappa
\partial_\rho \Phi -\lambda g'U \label{fi-inner},\\
-\varepsilon v\partial_\rho U &=& \partial^2_\rho U + \varepsilon
\kappa \partial_\rho U -\frac{\varepsilon v}{2}\partial_\rho
h(\Phi), \label{u-inner}
\end{eqnarray}
where $\kappa = \kappa_s +\kappa_w$ being $\kappa_s$ and $\kappa_w$
the two principal curvatures of the level-set $\phi=0$.
Inserting the inner expansion
\begin{eqnarray}
\Phi&=&\Phi_0+\varepsilon\Phi_1+O(\varepsilon^2)\\
U&=&U_0+\varepsilon U_1+O(\varepsilon^2) \label{eq:inner}
\end{eqnarray}
into Eqs. (\ref{fi-inner}) and (\ref{u-inner}) we
get, at zero order
\begin{eqnarray}
&&\partial^2_\rho \Phi_0 - f'(\Phi_0)-\lambda g'(\Phi_0)U_0=0
\label{inner-phi-0}\\
&&\partial^2_\rho U_0=0. \label{inner-u-0}
\end{eqnarray}
A valid solution of Eqs. (\ref{inner-phi-0}, \ref{inner-u-0})
is given by a kink for the phase-field
\begin{equation}
\Phi_0(\rho)=-\tanh (\frac{\rho}{\sqrt{2}}),
\end{equation}
and a vanishing $u$ at zero order, i.e. $U_0=0$.
At first order we get
\begin{eqnarray}
&&\Omega \Phi_1=-(v\alpha+\kappa)\partial_\rho \Phi_0 +\lambda
g'(\Phi_0)U_1
\label{eq:inner-phi-1}\\
&&\partial^2_\rho U_1 -\frac{\varepsilon v}{2}\partial_\rho
h(\Phi_0) = 0, \label{eq:inner-u-1}
\end{eqnarray}
where $\Omega$ is the linear self-adjoint operator given by
\begin{equation}
\Omega = \partial^2_\rho- f''(\Phi_0).
\end{equation}
Integrating Eq. (\ref{eq:inner-u-1}) over $\rho$ we obtain
\begin{equation}
\partial_\rho U_1 = A + \frac{v}{2} h(\Phi_0),
\label{eq:du1}
\end{equation}
where A is an integration constant independent of $\rho$. Integrating
Eq. (\ref{eq:du1}) over $\rho$ we get an expression for the diffusion field
at first order
\begin{equation}
U_1= B + A\rho +\frac{v}{2}\int_{0}^{\rho} d\rho' h(\Phi_0),
\label{eq:u1}
\end{equation}
where B is another integration constant.
To determine the value of the constant B we impose the solvability condition
for the existence of a solution for $\Phi_1$, which can be written as
\begin{equation}
\int_{-\infty}^{\infty} d\rho \;(\partial_\rho \Phi_0)
\;\Omega\Phi_1 = 0 \label{eq:solv}
\end{equation}
and allow us to find an expression for $B$
\begin{equation}
B=-\frac{(v\alpha+\kappa)I_1}{I_2\lambda}+\frac{v I_3}{2 I_2},
\label{eq:B}
\end{equation}
where the integrals $I_1$, $I_2$ and $I_3$ depend on the
particular choice of the functions $g$ an $h$ and are given by
\begin{eqnarray}
I_1&=&\int_{-\infty}^{\infty} d\rho  (\partial_\rho \Phi_0)^2 =
\frac{2\sqrt{2}}{3},
\label{eq:I_1}\\
I_2&=&-\int_{-\infty}^{\infty} d\rho (\partial_\rho \Phi_0)
g'(\Phi_0),
\label{eq:I_2}\\
I_3&=&\int_{-\infty}^{\infty} d\rho (\partial_\rho \Phi_0)
g'(\Phi_0), \int_{0}^{\rho} d\rho' h(\Phi_0). \label{eq:I_3}
\end{eqnarray}
In order to determine the value of the integration constant A, we
impose the matching of the inner and outer diffusion gradients
at the interface. This procedure will bring to an expression for
the conservation of the diffusion field across the interface. The
matching conditions for the derivatives of $u$ can be written in
the form
\begin{equation}
\lim_{r\rightarrow 0^{\pm}} \partial_r u({\bf r},t) = \lim_{\rho
\rightarrow \pm \infty} \frac{1}{\varepsilon}\partial_\rho
U(\rho,s,w,t). \label{eq:match-dru}
\end{equation}
We insert then the $\varepsilon$ expansions into Eq. (\ref{eq:match-dru})
and equal the corresponding orders.
For the matching of the derivatives at
zero order we get
\begin{equation}
\partial_ru_0|^\pm = A\pm \frac{v}{2},
\label{eq:match-dru0}
\end{equation}
where we have used Eq. (\ref{eq:du1}) and
that $h(\Phi_0(\rho \rightarrow \pm \infty))=\mp 1$,
which is the only restriction we have required for the
function $h$. This last equation can be used to find
a value for the constant $A$
\begin{equation}
A=\pm\frac{v}{2}+\partial_r u_0|^\pm. \label{eq:A}
\end{equation}
This last equation can be used to get an expression for the
conservation of the diffusive field across the interface
\begin{equation}
v = \partial_ru_0|^{-} - \partial_ru_0|^{+},
\label{eq:conservation}
\end{equation}
which is the form of the heat flux conservation across the
interface appearing in the sharp-interface problem Eq.
(\ref{si-cons}).

 The last thing we have to do is to recover the
Gibbs-Thomson equation in the thin-interface limit of the model.
For this we impose the matching conditions for the diffusion field
\begin{equation}
\lim_{r\rightarrow 0^{\pm}} u({\bf r},t) = \lim_{\rho \rightarrow
\pm \infty} U(\rho,s,w,t), \label{eq:match-u}
\end{equation}
and using Eqs. (\ref{eq:u1}, \ref{eq:B}) and (\ref{eq:A}) we obtain
\begin{equation}
u_{int}=-\frac{\varepsilon I_1}{I_2\lambda}\kappa
-\varepsilon\left[\frac{\alpha I_1}{\lambda I_2}-
\frac{I_3}{2I_2}-\frac{I_4}{2} \right] v, \label{eq:uint3}
\end{equation}
where the integral $I_4$ is given by
\begin{equation}
I_4 = \int_{0}^{\pm\infty} d\rho \left[ h(\Phi_0) \pm 1 \right].
\label{eq:I_4}
\end{equation}
Equation (\ref{eq:uint3}) can be directly compared with the
Gibbs-Thomson equation in the sharp-interface description (Eq.
\ref{si-gt}). Identifying terms we can get an expression for the
phase-field parameters in terms of the sharp-interface physical
constants
\begin{eqnarray}
&&d_0=\frac{W I_1}{\lambda I_2}
\label{eq:d0}\\
&&\beta=\frac{W}{D}\left[\frac{\alpha I_1}{\lambda I_2}-
\frac{I_3}{2 I_2}-\frac{I_4}{2} \right]. \label{eq:beta}
\end{eqnarray}
Note that in the last equations the integrals $I_{2}-I_{4}$ depend
on the particular choice for the functions $g$ and $h$. Another
interesting feature of this result is that the expression for
$\beta$ allows to simulate situations with neglecting kinetic
attachment by means of an appropriate selection of the phase-field
parameters.

\section{A second example: The Saffman--Taylor problem.}
\label{viscousfingering}

The so-called Saffman--Taylor or viscous fingering problem has played 
a central role in
the context of interfacial pattern formation because of its
relative simplicity both experimentally and in its theoretical
formulation \cite{general,st,bensimon,mccloud,couder,tanveer}. It deals with the 
destabilization
of the interface between two immiscible fluids when a less viscous
fluid is injected to displace a more viscous one, and / or when a
more dense fluid is placed on top of a less dense one in a
two-dimensional geometry known as a Hele-Shaw cell. This consists
of two parallel glass plates of dimensions $L_x \times L_y$
separated a distance $b \ll  L_x, L_y$.

In contrast with the three-dimensional Rayleigh--Taylor instability,
in such a cell and for some range of injection velocities $V_{\infty}$,
the flow is overdamped (small Reynolds number limit). Therefore,
the velocity $\vec u$ is proportional to the gradient of a potential 
defined as the deviation from hydrostatic 
pressure: 
$\vec u = -(b^2/12\mu) \vec\nabla (p-\rho g_{\rm eff} y)$, where $p$ is the 
pressure,
$\mu$ the shear viscosity, $\rho$ the density and $g_{\rm eff}$ an effective
gravitational acceleration in the $y$ direction for non-horizontal cells.

\subsection{The sharp-interface model}
\label{fbpvf}

Assuming incompressibility in the bulk, $\vec\nabla\cdot \vec u=0$, 
continuity of normal velocities and a pressure drop proportional to the
interfacial
tension $\sigma$ on the interface, one obtains the classical free boundary
problem
\begin{eqnarray}
\label{laplace} \nabla^2 p &=& 0,
\\
\label{normalvelocities} \frac{\hat r}{\mu_1}\cdot \left (
\vec\nabla p_1+ \rho_1 g_{\rm eff} \hat y \right) = \frac{\hat
r}{\mu_2}\cdot \left ( \vec\nabla p_2+ \rho_2 g_{\rm eff} \hat y
\right) &=& -\frac{12}{b^2}v_n,
\\
\label{pressuredrop} p_1-p_2 &=& \sigma\kappa,
\end{eqnarray}
where the first equation holds in the bulk of each fluid, and the next are its
boundary
conditions on the interface between them;
subscripts $1$ and $2$ label each fluid and the notation $p_i$
stands for the pressure on the interface coming from fluid $i$;
$r$ is a coordinate normal to the interface, $v_n$, its normal velocity, and
$\kappa$, its curvature.

The dynamics are controlled by the two dimensionless
parameters
\begin{equation}
\label{eq:bandc}
B=\frac{b^2\sigma}{12L_x^2[V_\infty(\mu_1-\mu_2)+g_{eff}(\rho_1-
\rho_2)]}>0, \;\;\;\; c=\frac{\mu_1-\mu_2}{\mu_1+\mu_2}>0.
\end{equation}
$B$ is a dimensionless surface tension,
and can be understood as the ratio between the capillary (stabilizing) force
and the
driving (destabilizing) force (injection+ gravity), and $c$ is the viscosity
contrast, which is so far completely arbitrary: $0\le c\le 1$.
This corresponds to having set ourselves in the frame moving with the fluid
at infinity (or, equivalently, with the mean interface) and taken $L_x$ as unit
length
and $U_*\equiv cV_\infty+g\frac{\Delta\rho}{\mu_1+\mu_2}$ as unit velocity
(see Ref. \cite{aref1}).

Analytical understanding of these highly nonlinear and nonlocal
dynamics [Eqs. (\ref{laplace}--\ref{pressuredrop})] is basically
restricted to high viscosity contrast $c=1$ and small surface
tension \cite{Siegel,magdaleno00,paune02}, so one relies mostly on
numerical work
\cite{aref1,paune02,aurora,jaume,vinals,shelley,shelleyreview}.
A systematic weekly nonlinear analysis is also possible for the
early stages of the evolution and arbitrary parameters
\cite{alvarez01}. In the next subsection we introduce a PFM for
Hele-Shaw flows with arbitrary viscosity contrast $c$. Although in
the high contrast limit $c=1$ the Hele-Shaw dynamics are quite
analogous to the one-sided solidification problem (in the
appropriate approximations \cite{vinals}), the arbitrary viscosity
contrast case has been shown to exhibit quite different dynamics,
and has raised some interesting questions, particularly concerning
the sensitivity of finger competition to viscosity contrast
\cite{aref1,aurora,jaume,maher} and the long time
asymptotics of the low viscosity contrast limit $c=0$
\cite{jaume,paunethesis}.

\subsection{Discussion of the sharp interface limit}
\label{secpf}

As emphasized in the introduction, the classical Saffman--Taylor problem does
not involve phase transformations, so that a derivation of a PFM for it from a
free
energy functional is not necessary. Although such a variational model is possible
\cite{variationalvf}, here we discuss the first PFM for the Saffman--Taylor problem
with arbitrary viscosity contrast of Ref. \cite{vf1},
without the help of such a functional, as done in that reference,
in order to illustrate how such a derivation (and the subsequent sharp-interface
limit) work for a particular case.
The idea is to construct two coupled partial differential equations,
one for some scalar field describing the flow and one for a phase field,
{\em such} that we recover Eqs. (\ref{laplace}--\ref{pressuredrop})
in the sharp-interface limit. How this can be achieved depends of course of the
particular free boundary problem.

An important difference between the arbitrary $c$ Saffman--Taylor
problem Eqs. (\ref{laplace}--\ref{pressuredrop}) and the
solidification one, as stated by Eqs. (\ref{diff}--\ref{gibbs0}),
lies in their boundary conditions: in the two-sided solidification
problem the physical field $u$ is continuous and its normal
gradients, discontinuous, through the interface; in the
Saffman--Taylor problem it is the other way round, a velocity
potential is discontinuous and its normal gradients are continuous
(because the velocities are so). We therefore begin by rewriting
the free boundary problem in terms of the harmonic conjugate of
the velocity potential, the stream function $\psi$. This exchanges
normal and tangential derivatives ($\hat x\cdot\vec u=\partial_y
\psi$, $\hat y\cdot\vec u=-\partial_x \psi$) so that we obtain, in
the adimensionalisation mentioned in the last section,
\begin{eqnarray}
\label{eq:laplace}
\nabla^2\psi&=&0,\\
\label{eq:discontinuity}
\psi_r(0^+)-\psi_r(0^-)&=&-\gamma-c[\psi_r(0^+)+\psi_r(0^-)],\\
\label{eq:continuity} \psi_s(0^+)=\psi_s(0^-)&=&-v_n,
\end{eqnarray}
where $s$ is arclength along the interface,
subscripts stand for partial derivatives except for $v_n$, and
\begin{equation}
\label{eq:gammasharp} \frac{\gamma (s)}{2}\equiv B\kappa_s+\hat y
\cdot \hat s,
\end{equation}
with $\kappa (s)$ the interface curvature.

As we can see, the boundary conditions continue to be substantially
different from those of two-sided solidification, 
since the stream function $\psi$
is now continuous, but the value it takes on the interface and the magnitude
of its normal derivative jump are not the same than in solidification. 
Nevertheless, because $\psi$ is continuous through
the interface, a single field (and not one field in each fluid) can be used
for the whole bulk, which is obviously advantageous when it comes
to build a PFM, since we recall that PFM treat all the system as bulk.

Note that Eqs. (\ref{eq:laplace},\ref{eq:discontinuity}) can be written
together as
\begin{equation}
\label{eq:poisson} \nabla^2\psi=-w, \;\;\;\; w=\{\gamma
(s)+c[\psi_r(0^+)+\psi_r(0^-)]\}\delta(r)
\end{equation}
where $\delta(r)$ is the Dirac delta distribution and $w\equiv
\hat z\cdot (\vec\nabla \times\vec u)$ is the fluid vorticity,
which is confined to the interface.
This suggests to construct our PFM equation for $\psi$
as a regularisation of this distribution
equality, coupled to the remaining boundary condition, the continuity of normal
velocities, by advecting the phase field $\theta$ (and therefore the interface)
precisely with the physical fluid velocity on the interface.

More specifically, the equations proposed in Ref. \cite{vf1} read
\begin{eqnarray}
\label{eq:sf} \epsilon \frac{\partial\psi}{\partial
t}&=&\nabla^2\psi+c\vec \nabla \cdot (\theta \vec \nabla
\psi)+\frac{1}{\epsilon} \frac{1}{2\sqrt 2}
\gamma(\theta )(1-\theta^2),\\
\label{eq:pf} \epsilon^2 \frac{\partial \theta}{\partial
t}&=&f(\theta)+\epsilon^2\nabla^2\theta +\epsilon^2 \kappa(\theta
) |\vec \nabla \theta |+\epsilon^2 \hat z \cdot (\vec \nabla \psi
\times \vec \nabla \theta)
\end{eqnarray}
where $f(\theta )\equiv \theta (1-\theta^2)$, and
$\frac{\gamma(\theta)}{2}\equiv \hat s(\theta)\cdot(B\vec\nabla
\kappa(\theta) +\hat y)$, $\kappa(\theta)\equiv -\vec\nabla \cdot
\hat r(\theta)$,
with
$\hat r(\theta)\equiv \frac{\vec\nabla \theta}{|\vec\nabla \theta|}$
and $\hat s(\theta)\equiv \hat r(\theta) \times \hat z$.
$\gamma(\theta)$,
$\kappa(\theta)$ are functionals which
generalize the magnitudes defined above for the interface to
any level-set of the phase-field, and
$\epsilon$ is a small parameter of the model;
in particular it can represent the interface thickness $W$ (see below).
This model is inspired in the vortex-sheet
formulation of the problem \cite{aref1}.
Similar ideas have previously been
applied to describe physically diffuse interfaces in the
context of steady state selection in thermal plumes \cite{benamar}.

If we leave the two last terms aside, Eq. (\ref{eq:pf}) is the time-dependent
Ginzburg-Landau 
equation for a non-conserved order parameter or
model A (without noise) in the classification of Ref. \cite{halperin:74}. 
The field in this model is
known to relax towards a kink solution in a
short time scale, and then to evolve to minimize the length of the
effective interface according to the Allen--Cahn law (i.e. with normal
velocity proportional to the local curvature). The factor
multiplying the laplacian has been chosen to be $\epsilon^2$ for
the kink width to be ${\cal O}(\epsilon)$, so that $\epsilon$ can
be considered the interface thickness.
On the other hand, the $\epsilon^2$ factor in the time derivative
ensures that the relaxation towards the kink is much faster than
the evolution of the interface. Notice that model A describes the
relaxational dynamics of a non-conserved order parameter, whereas
our problem is actually non-relaxational and strictly conserved
(mass conservation and immiscibility). The other two terms in the
phase-field equation will correct this apparent contradiction. In
order to cancel out the local Allen-Cahn dynamics of the interface
which is built in model A, we add the term $\epsilon^2
\kappa(\theta ) |\vec \nabla \theta |$. Such a
term cancels out Allen-Cahn law by giving rise, to leading order,
to an identical contribution but with opposite sign.

With these elements so far, our phase-field relaxes to a kink
profile located along an arbitrary interface which remains almost
completely stationary, regardless of its shape, provided that the
interface thickness remains smaller than the local radius of
curvature (i.e., that a sharp-interface description makes sense).
This is because the dynamical effect of surface tension associated
to the Ginzburg-Landau free energy of model A without noise has
been removed (up to first order) and the interface has not yet
been coupled to the fluid flow, represented by the stream
function. This coupling is achieved by adding the last term in Eq.
(\ref{eq:pf}), which stands for $-\epsilon^2 \vec u\cdot \vec
\nabla \theta$ and thus sets the phase-field
---and therefore the interface--- in the frame moving with the fluid velocity
$\vec u$. This term restores the fully nonlocal dynamics of the Hele--Shaw
model. In particular it yields the continuity of normal velocities
Eq. (\ref{eq:continuity}) and reintroduces surface tension, which is
contained in the dynamical equation for the stream function through
$\gamma(\theta)$.

At first sight, the idea presented in Ref. \cite{vf1} and reviewed
here of cancelling out some undesired effect (here, the Allen-Cahn
law) from the PFM equations by adding an extra term designed to
produce the same effect but of opposite sign close to the
sharp-interface limit [here, the term  $\epsilon^2\kappa(\theta)
|\vec \nabla \theta |$] might seem somewhat artificial or
``unphysical'' compared to PFM of solidification where the surface
tension arises more ``naturally'' from the free energy cost for
the gradients of the phase field without the need to cancel it out
and reintroduce it in a different way. Nevertheless, it has been
successful in later developments, both in the sharp-interface
limit and at first order in the interface thickness. 
The present formulation has recently been
adapted to treat vesicles \cite{vesicles}. Seen as an interface,
the vesicle membrane not only evolves to conserve mass inside and
outside the vesicule, but also keeps its own length roughly
constant. Consequently, another term was added to cancel out
interface length changes, yielding a suitable model to treat such
membranes \cite{vesicles}. Similarly, an extra term in the
standard PFM of solidification cancels out a spurious solute
trapping correction to the the one-sided model at first order in
the interface thickness \cite{short1sided,long1sided}. Combined
with an appropriate choice of the model functions, all first-order
corrections can then be cancelled out \cite{short1sided,long1sided}.

As for Eq. (\ref{eq:sf}), its right hand side is intended to
reproduce Eq. (\ref{eq:poisson}), and therefore also Eqs.
(\ref{eq:laplace}) and (\ref{eq:discontinuity}). If the
phase-field $\theta$ has a kink shape, $1-\theta^2$ is a peaked
function which, when divided by $\epsilon$, gives rise to the
delta distribution for the vorticity. However, this only accounts
for the $\gamma$ in the weight of the delta. The part proportional
to the viscosity contrast $c$ must be introduced separately as the
$c\vec \nabla \cdot (\theta \vec \nabla \psi)$ term because of the
non-local character of $\psi_r(0^+)+\psi_r(0^-)$. Finally, the
time derivative is multiplied by $\epsilon$ to recover the
laplacian (and not diffusive) behavior of Hele--Shaw flow in the
sharp-interface limit.

In the PFM the interface width and the convergence to the sharp
interface limit is controlled by the small but finite value of the parameter
$\epsilon$.
Note that the role of the sharp-interface limit in this PFM is not the same
than in solidification, in the sense that the parameters of the model do not
need to be identified, but are just ``built in'' in the model.
The sharp-interface limit hence merely serves as a safety check.
Nevertheless, one can still wonder
which value of $\epsilon$ is needed to accurately reproduce
the actual Hele--Shaw dynamics for given values of the physical parameters $B$
and $c$.
This question can be qualitatively answered by noting the distinct roles
played by $\epsilon$ in the phase-field equations, Eqs. (\ref{eq:sf},\ref{eq:pf}):

The $\epsilon$ factors appearing in $\epsilon^2\nabla^2\theta$,
$\epsilon^2 \kappa(\theta ) |\vec \nabla \theta |$ and
$\frac{1}{\epsilon} \frac{1}{2\sqrt 2} \gamma(\theta
)(1-\theta^2)$ all stand for the interface thickness, and this is
required to be small compared to the longitudinal length scale
$|k|^{-1}$ of the interface: $\epsilon |k|<<1$.

In contrast, the $\epsilon$ in $\epsilon
\frac{\partial\psi}{\partial t}$ has nothing to do with the
interface thickness (and we will therefore denote it by
$\tilde{\epsilon}$ from now on), but its aim is to ensure that the
stream function is laplacian and not diffusive in the
$\tilde{\epsilon} \rightarrow 0$ limit, which commutes with the
$\epsilon \rightarrow 0$ one: $\tilde{\epsilon}$ sets the time
scale of the diffusion of the stream function through a given
characteristic length of wavenumber $k$,
$\frac{\tilde{\epsilon}}{(1\pm c)k^2}$, which must be much smaller
than the characteristic growth rate of the interface
$|\omega|^{-1}$, so that the stream function is slaved to the
interface: $\frac{\tilde{\epsilon}|\omega|}{k^2}<<1\pm c$.

The $\epsilon^2$ in
$\epsilon^2 \frac{\partial \theta}{\partial t}$ represents the
relaxation time of the phase field towards the steady kink solution (see Eq.
\ref{eq:pf}),
which must be kept well
below the interface growth time $|\omega|^{-1}$ for the phase-field to remain
close to the kink profile during the interface evolution:
$\epsilon^2|\omega|<<1$. This factor must be the same that the one in
$\epsilon^2 \hat z \cdot (\vec \nabla \psi \times \vec \nabla
\theta)$ in order to recover the macroscopic
equation Eq. (\ref{eq:continuity}).
In fact  there are at least two distinct powers of $\epsilon$ for this relaxation
time ($\epsilon$ and $\epsilon^2$) for which the right sharp-interface limit is
achieved, and the corrections to it are also the same.

     To sum up, there are at least two independent small parameters
($\epsilon$ and $\tilde{\epsilon}$)
controlling the limit. When trying to approach macroscopic solutions
by means of numerical
integration of the phase-field equations,
it is very convenient to vary them independently in order to save computing
time, since both affect it \cite{vf2}.

A more quantitative answer to the question of the necessary values of
$\epsilon$, $\tilde{\epsilon}$ to achieve a given precision can be obtained by
computing not only the sharp-interface limit, but also the corrections to it at
first order
in the interface thickness $\epsilon$ considering $\tilde{\epsilon}$ of
${\cal O}(\epsilon)$.
Thus, one obtains an effective free boundary problem (where the interface is
mathematically sharp)
keeping track of corrections partly due to the fact that the interface
of the PFM is {\em not} sharp. More precisely, it keeps track of corrections
up to first order in both $\epsilon$ and $\tilde{\epsilon}$.
This limit follows basically the same procedure described in Sec.
\ref{thininterface}.
It is performed in Ref. \cite{vf1}, where a detailed account of how the successive
orders in the asymptotic expansions are derived and matched can be found.
Here, we just give the result for the sake of brevity:
\begin{eqnarray}
\label{eq:difusio} \tilde{\epsilon}\frac{\partial \psi}{\partial
t}&=&(1\pm c)\nabla^2\psi
\\
\label{eq:psir}
\psi_r(0^+)-\psi_r(0^-)&=&-\Gamma \\
\label{eq:psis} \psi_s(0^\pm)&=& -v_n -\epsilon\frac{\sqrt
2}{2}\frac{\Gamma_s}{2}g_\pm(c)
\;\;\;\;\nonumber \\
&=&-v_n+\epsilon{\sqrt 2}\frac{\Gamma_s}{2} [\frac{5}{6}\mp
c+\frac{2}{5}c^2+{\cal O}(c^3)],
\end{eqnarray}
where $\Gamma\equiv\gamma+c[\psi_r(0^+)+\psi_r(0^-)]$ is the weight of the
vorticity defined in Eq. (\ref{eq:poisson}) evaluated up to ${\cal O}(\epsilon)$
and
\begin{equation}
g_\pm(c)= 1-\frac{1}{c^2}+\left (\pm
1+\frac{\frac{1}{c^3}-\frac{3}{c}}{2}\right )\ln\frac{1+c}{1-c}.
\end{equation}

This contains two important results:
On the one hand, the correct free boundary problem
of Eqs. (\ref{eq:laplace}-\ref{eq:continuity}) is recovered in the sharp-interface
limit $\epsilon, \tilde{\epsilon} \rightarrow 0$, as expected.
On the other hand, the corrections at ${\cal O}(\epsilon,\tilde{\epsilon})$
in Eqs. (\ref{eq:difusio}) and (\ref{eq:psis})
are obtained and go as $\tilde{\epsilon}$ and $\epsilon$ respectively,
whereas Eq. (\ref{eq:discontinuity}) remains unaffected. Note that the correction
in $\epsilon$ appearing in Eq. (\ref{eq:psis})
has nothing to do with an Allen--Cahn law. So the
$\epsilon^2 \kappa(\theta ) |\vec \nabla \theta |$ term
has cancelled that effect out even in the first order corrections.
This is the partial improvement of the convergence to the sharp-interface
limit achieved by this model and mentioned in the introduction.

\section{Applications}

In this Section we present results obtained in simulations of PFM in different
pattern forming systems. We will start with two examples taken from solidification
problems, namely directional solidification of alloys, and dendritic sidebranching
in free solidification.
In these problems we will devote special attention to the effect of fluctuations.
Next we will apply the same kind of models to the growth of mesophases in liquid
crystal. This problem is mathematically similar to those of solidification, but
with the particularity of the presence of strong anisotropies, both in the
interface and in the bulk. We will show how the PFM is able to deal these
anisotropies and to reproduce the morphological transitions observed in
experiments. Finally, we present some results for viscous fingering obtained with
the PFM discussed above.

\subsection{Directional solidification}

 The first application will be the study of initial transients in directional
solidification of an alloy in the context of the symmetric model 
\cite{langerrmp}.
This example will be used to show the numerical convergence of the
PFM to the sharp interface model. Once the convergence is reached,
the PFM will be used to quantitative test of theoretical
predictions. This will be done in the context of the symmetric
model of solidification. To that end we will present predictions
for the transient recoil of the sharp interface model and for the
transient dispersion relation by using an adiabatic approximation.
We will then perform phase field simulations of this initial
regime, both for the evolution of single modes and for the
complete system with thermal fluctuations.

The selection of a dendritic pattern during the directional
solidification of a dilute binary alloy is a complex problem which
depends on initial conditions\cite{WL90}, and in particular on the
first wavelengths that appear in the destabilization of the planar
front induced by fluctuations \cite{WL,car2}. Recent work has
focussed in the importance of internal fluctuations in
solidification patterns \cite{brener95,pocheau}. Quantitative
agreement with experiments has only been obtained for the
solidification of pure substances \cite{cummins}, whereas in
solutal growth the origin itself of the fluctuations is still an
open problem \cite{ricard}. The analysis of the evolution of
single modes and of fronts with fluctuations, both from theory and
from simulations, should be of relevance in the dendritic
selection problem.

\subsubsection{Predictions of the sharp-interface model}
In a directional solidification experiment, a thermal gradient
$G\hat{\bf z}$ is moved
in the $\tilde{z}$ direction along the sample at constant pulling velocity
$V\hat{\bf z}$. Provided the sample is thin, we take the system as 2D and describe
the
interface position in the moving frame of the gradient by
$\tilde{z}=\tilde{\xi}(\tilde{\vec{\rho}},\tilde{t})$,
where $\tilde{\vec{\rho}}=\tilde{x}\hat{\bf x}+\tilde{y}\hat{\bf y}$.
The solid phase is located in the
region where $\tilde{z}<\tilde{\xi}$, and the liquid
where $\tilde{z}>\tilde{\xi}$.
We will consider the particular case of symmetric directional
solidification, which assumes the same solute diffusivity $D$ in both
phases.
We introduce diffusion length $l=D/V$ and time $\tau=D/V^{2}$ to scale variables
as $\vec{r}=\vec{\tilde{r}}/l$ and $t=\tilde{t}/\tau$.
We also introduce a diffusive field in each phase
$u_{i}(\vec{r},t)=\frac{C_{i}- C_{\infty}}{\Delta C_{0}}$ (i=1
solid, i=2 liquid), where
$C_{i}$ is the solute concentration,
$C_{\infty}$ the concentration
far away from the solid-liquid interface,
and $\Delta C_{0}=[C_{2}-C_{1}]_{int}$ the concentration jump across the
interface.
In the moving frame and in reduced variables, the fields $u_{i}$
evolve according to the diffusion equation
\begin{equation}
(\frac{\partial}{\partial t} - \frac{\partial}{\partial z} -
\nabla^{2}) u_{i}(\vec{r},t)=0. \label{eq:1}
\end{equation}

The diffusion fields $u_{i}$ must satisfy some moving boundary
conditions
at the interface position which can be written as
\begin{eqnarray}
&[u_{1}-u_{2}]_{int}=-1 \label{eq:2}
\\
&u_{2}|_{int}=1-\frac{l}{l_{T}}\xi+\frac{d_{0}}{l}\kappa
\label{eq:3}
\\
&{\bf \hat{n}} \cdot \big[ \vec{\nabla} u_{1} - \vec{\nabla} u_{2}
\big]_{int}= -n_{z} (1+\dot{\xi}) \label{eq:4}
\end{eqnarray}

Eqs. (\ref{eq:1}-\ref{eq:4}) define the so called sharp-interface
description of the symmetric directional solidification problem.
Eq. (\ref{eq:2}) relates solute concentrations at both sides
of the interface. We have assumed the additional approximation of having
a constant concentration jump across the interface. This particular
assumption is equivalent to suppose that the mixture has parallel
solid and liquid branches in the $T(C)$ coexistence diagram, which is valid
only for liquid crystals and alloys with
a partition coefficient close to $1$.
Eq. (\ref{eq:3}) is the Gibbs-Thomson equation
(local equilibrium at the interface).
In this relation
$l_{T}=|m_{L}|\Delta C_{0}/G$ is a thermal length
imposed by the temperature gradient, and $d_{0}$ is the capillary length.
Eq. (\ref{eq:4}) describes the solute conservation across the
interface, and
$\vec{\hat{n}}$ represents a normal unitary vector
pointing to the liquid.

Using Green's function techniques \cite{car1}, it is possible to
derive a closed integral expression for $u_i$ at each side of the
interface. Introducing the notation $p=(\vec{\rho},z,t)$,
$p_{S}=(\vec{\rho},\xi,t)$, the corresponding integro-differential
equation reads
\begin{eqnarray}
\frac{l}{l_{T}}\xi(\vec{\rho},t_{0}) =
&-\int_{-\infty}^{\xi(\vec{\rho},t_{0})} d\vec{r'}
G(p_{s};\vec{r'},t_{0}) u_{1}(\vec{r'},t_{0})-
\int_{\xi(\vec{\rho},t_{0})}^{\infty} d\vec{r'}
G(p_{s};\vec{r'},t_{0}) u_{2}(\vec{r'},t_{0})+ \nonumber
\\
&\int_{t_{0}}^{t} dt' \int d\vec{\rho'}
(1+\dot{\xi}(\vec{\rho}',t)) G(p_{s},p_{s}')
+\frac{1}{2}(1+erf(\frac{\xi(\vec{\rho},t)-\xi(\vec{\rho},t_{0})+t}
{2\sqrt{t}})) \label{eq:7}
\end{eqnarray}
where $G(p,p')$ is the Green function for Eq. (\ref{eq:1}), and
$\xi=0$ stands for the steady position of an unperturbed planar
interface. Note that Eq. (\ref{eq:7}) includes transients from the
initial condition at $t_{0}$.
The next step is to perform a linear stability analysis
of the problem to obtain a transient dispersion relation
describing the time evolution of a sinusoidal modulation with wavevector $kl$.
Within an adiabatic approximation \cite{WL,car2} we derive the growth ratio of the
mode as
\begin{eqnarray}
\omega(kl,t)=\frac{l}{l_{T}}\dot{\xi} -(1+\dot{\xi})^{2} +q_{2}
(1+\dot{\xi})
-(q_{1}+q_{2})\left[\frac{l}{l_{T}}+\frac{d_{0}}{l}(kl)^{2}\right],
\label{eq:8}
\end{eqnarray}
where
\begin{eqnarray}
&q_{\alpha}=\frac{(-1)^{\alpha}}{2}(1+\dot{\xi})+\sqrt{\frac{1}{4}(1+\dot{\xi}
)^{2}+(kl)^{2}+\omega(kl,t)}. \label{eq:8bis}
\end{eqnarray}
Therefore the prediction of the transient dispersion relation consists of two
steps. Firstly the numerical resolution of Eq. (\ref{eq:7}) for a planar interface
is performed by a Newton-Raphson method, yielding the transient front
position $\xi(t)$. This function is then introduced into Eq. (\ref{eq:8}) to
obtain $\omega(kl,t)$ in this adiabatic approximation.

\subsubsection{Results from the phase-field model}

We will use a PFM introduced by Losert {\it et al.} \cite{Karma1}
for symmetric directional alloy solidification with constant
miscibility gap. This model can be obtained from that of Sec.
\ref{thininterface} by making $\beta = 0$ ({\it i.e.} no kinetic
dynamics) in Eq. \ref{eq:beta}. Once the thermal gradient is
included, the equations for the phase field and the concentration
take the following form in reduced variables:
\begin{eqnarray}
&\frac{a_{1}a_{2}}{d_{0}/l} \varepsilon^{3} \partial_{t}\phi =
\varepsilon^{2} \nabla^{2}\phi + (1-\phi^{2}) \large( \phi -
\varepsilon\frac{a_{1}}{d_{0}/l}(1-\phi^{2})
(u+\frac{z-t}{l_{T}/l}) \large) \label{eq:9}
\\
&\partial_{t} u = \nabla^{2} u +\frac{1}{2}\partial_{t}\phi.
\label{eq:10}
\end{eqnarray}
$a_{1}=\frac{5\sqrt{2}}{8}$ and $a_{2}=0.6267$ are integral
constants obtained when performing the thin-interface limit
\cite{karma:98}. Note that the equation for the phase field
evolution contains three parameters: two main control parameters
coming from the sharp-interface model ($l_{T}/l$ and $d_{0}/l$),
and the model-specific parameter $\varepsilon$.

We will consider the transient from the rest, {i.e.} for an initial condition at
$t_{0}=0$ consisting of an equilibrium solid-liquid planar interface
located at $\xi(t=0)=l_{T}/l$. In this case
$C_{1}(\vec{r},0)=C_{\infty}-\Delta C_{0}$ ($u_{1}(\vec{r},0)=-1$), and
$C_{2}(\vec{r},0)=C_{\infty}$ ($u_{2}(\vec{r},0)=0$).That corresponds
to taking $u(\vec{r},0)=-1$ as initial condition for the phase-field model.
Trying to mimic a real experiment, we consider here that at
$t=0$ the pulling velocity suddenly takes the final value V (1 in scaled
variables).

We perform numerical integration of the PFM Eqs.
(\ref{eq:9},\ref{eq:10})
with an explicit finite-differences scheme with $\Delta x=0.8$ and
$\Delta t=0.08$ .
We first study the 1D dynamical evolution during the transient,
comparing the results with the sharp-interface predictions. Fig. 1 presents the
front position for two different values of the control
parameter $l_{T}/l=12.5$ and $2.5$.
For each case, we compare
simulations for three different values of $\varepsilon$
($\varepsilon=0.5$, $\varepsilon=0.25$ and $\varepsilon=0.125$) with the
front position obtained with direct resolution of the integral equation
(\ref{eq:7}). Convergence to the sharp-interface limit can be observed as
$\varepsilon$ decreases, and good agreement is found for a
value of $\varepsilon=0.125$.
\begin{figure}
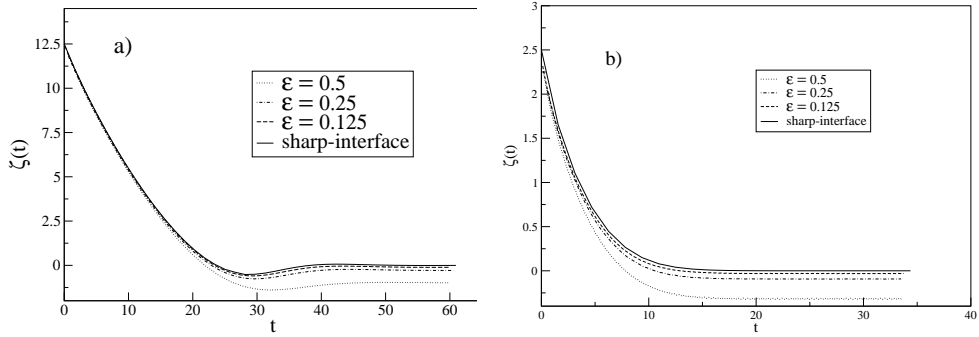

\includegraphics[width=.5\textwidth]{psi12.5.eps}
\hspace{0.1cm}
\includegraphics[width=.5\textwidth]{psi2.5.eps}
\caption{Evolution of the interface position during the transient
for (a) $l_{T}/l=12.5$ and (b) $l_{T}/l=2.5$, and convergence to
the sharp-interface as $\varepsilon \rightarrow 0$.}
\end{figure}

We now estimate the transient dispersion relation $\omega(kl,t)$
from PFM simulations and compare it with the sharp interface
prediction of Eqs. (\ref{eq:8}, \ref{eq:8bis}). To this end we
simulate for each desired value of $t$ a (planar) 1D interface
evolving from $t_0=0$ to $t$. At that moment we introduce a
sinusoidal interface perturbation with wavevector $kl$, and
continue the simulation in 2D. The spectral analysis of the front
allows us to locate the regime where the mode evolution is linear.
This is represented in Fig. 2a, where three definite regions can
be observed. From the linear region the value of the transient
growth rate $\omega(kl,t)$ is calculated. Fig. 2b shows the growth
rate at different times obtained for two different modes $kl=1.25$
and $kl=2.5$ in the case of $l_{T}/l=12.5$ and $d_{0}/l=0.06923$.
Quantitative agreement is observed for all times.
\begin{figure}
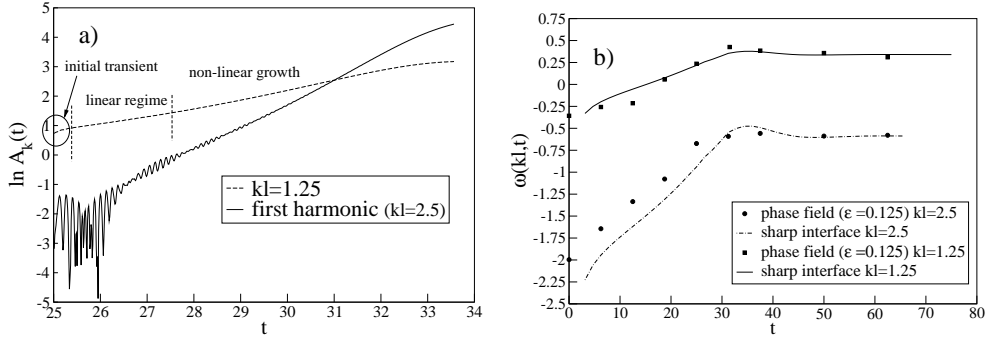

\begin{minipage}{\textwidth}
\includegraphics[width=.5\textwidth]{linear.eps}
\hspace{0.2cm}
\includegraphics[width=.5\textwidth]{omega.eps}
\caption{ a) Localization of the linear growth regime of the mode
$kl=1.25$. b) Time evolution of two different modes $kl=1.25$ and
$kl=2.5$ for $l_{T}/l=12.5$. }
\end{minipage}
\end{figure}

We can now introduce internal fluctuations in the diffusion
equation of the phase-field by introducing the noises appearing in
Eqs. (\ref{pf1}-\ref{q}). In simulations the growth of a range of
wavelengths can be observed until a selected mode dominates. The
spectral analysis of the front reveals quantitative agreement
during the linear regime between the growth of each mode and the
transient dispersion relation predicted by the sharp interface
model. In Fig. 3 we present the amplitude of one of the modes
($kl=2.5$, $l_{T}/l=12.5$) averaged for 15 different noise
realizations. There is a good agreement in the regime with
positive $\omega$ (i.e. were an increasing $A_k(t)$ is predicted).
\begin{figure}
\begin{center}
\includegraphics[width=7.1cm]{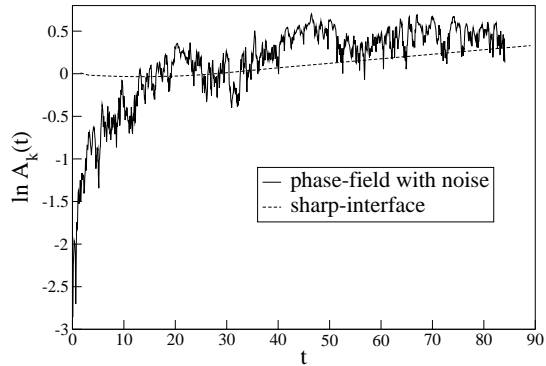}
\caption{Time evolution of the mode $kl=2.5$ ($l_{T}/l=12.5$) with
the phase-field with noise, and comparison with the
sharp-interface prediction.}
\end{center}
\end{figure}

\subsection{Dendritic sidebranching in free solidification}

In the previous Section our interest was the initial
destabilization of the planar interface in solidification
experiments, and the influence of the internal noise. That
transient has relevance in the selection of the wavelength of the
dendritic array. In the present Section our interest will
lie in some characteristics of the dendrite itself, in particular
in those of the sidebranching. We study the noise--induced
sidebranching scenario in the linear regime and in particular we
focus on the issue of the internal {\it vs.} external origin of
the noise \cite{ricard}.

In a frame of reference moving with the tip of the dendrite,
sidebranching can be seen as a wave that propagates along the
dendrite away from the tip at the tip's velocity. Therefore, an
appropriate characterization is provided by the amplitude of the
wave. Brener and Temkin \cite{brener95} pointed out that
experimentally observed sidebranching could be explained by
considering noise of thermal origin. They analytically found that
the growth of sidebranching amplitude in the linear regime behaves
exponentially as a function of $z^{2/5}$.
Dougherty {\it et al.} \cite{dougherty87} studied sidebranching in
experiments with ammonium bromide dendrites. The sidebranching
amplitude was found to increase exponentially up to a certain
value of $z$, from which the linear theory is presumably no longer
valid. It was also observed that side branches separated by more
than about six times the mean wavelength are uncorrelated.

Bisang and Bilgram \cite{bisang96} found a quantitative agreement
between the predictions for the linear regime in Ref.
\cite{brener95} and their results in experiments with xenon
dendrites. They concluded that Brener and Temkin calculations
describe correctly the sidebranching behaviour of dendrites for
any pure substance with cubic symmetry and thus thermal noise was
the origin of the sidebranching observed in their experiments.

Karma and Rappel \cite{karma:99} included thermal noise in a
two--dimensional PFM for solidification controlled by heat
diffusion, and obtained a good quantitative agreement between the
computed sidebranching amplitude and wavelength as a function of
distance to the tip and the predictions of the linear $WKB$ theory
for anisotropic crystals in two dimensions. For a needle crystal
shape $x \sim z^{3/5}$, sidebranching amplitude was found to
increase exponentially as a function of $z^{2/5}$.

However, the extension of the theory of dendritic sidebranching to
the growth controlled by solutal diffusion shows that there are
indications that in some experiments internal thermodynamical
fluctuations could not account for the observed sidebranching
activity \cite{ricard}. In that case, some other source of
fluctuations, of external origin, should be invoked. We study some
of the consequences derived from adding a non--conserved noise
source into a two--dimensional PFM. This noise is of very
different nature than what one should employ to account for
internal fluctuations.

\subsubsection{Sidebranching characteristics}

We have performed simulations of dendritic growth by employing a
PFM for solidification with anisotropy included in the surface
tension \cite{wang:93}, which has been taken as
$\sigma=\sigma(0)(1+\gamma_{su} cos(4\theta))$. Whereas internal
fluctuations would have appeared as a stochastic current (and,
therefore, as a conserved term) in the equation for the diffusion
field and an additional stochastic term in the phase field
equation , like in Eqs.
(\ref{variational1},\ref{variational2}), we have chosen to model
an external source of fluctuations by a non--conserved random term
added to the equation for $u$. This random term reads simply  $I
\cdot r$, where $I$ denotes the amplitude of the noise, and $r$ is
a uniform random number in the interval $[-0.5,0.5]$. More details
about the employed numerical procedure and the selection of
parameters can be found in Ref. \cite{ricard}.

\begin{figure}
\begin{center}
\includegraphics[width=6cm]{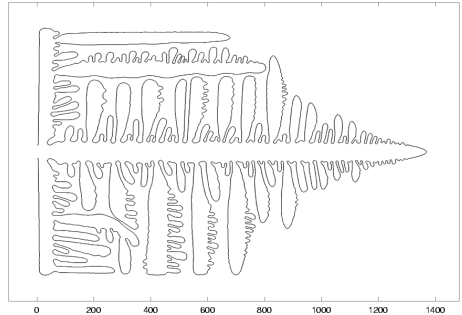}
\caption{\small Dendrite obtained with the PFM with an external
noise term. \label{fig42a} }
\end{center}
\end{figure}

\begin{figure}
\begin{center}
\includegraphics[width=6cm]{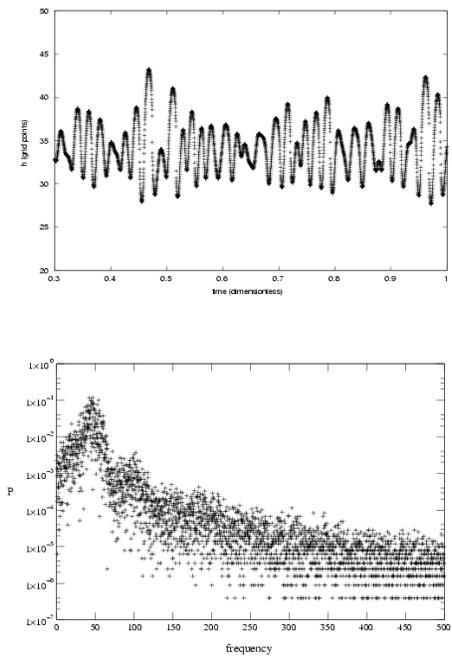}
\caption{\small Half--width $h_{z}(t)$ of the dendrite (up) and
its power spectrum $P_{z}(f)$ (down) for $z=100$.  \label{fig42b}}
\end{center}
\end{figure}

A growing dendrite obtained by means of the PFM is shown in Fig.
\ref{fig42a}. Side branches appear at both sides of the main
dendrite, yielding approximately a $90^o$ angle with it, like it
was observed in Ref. \cite{dougherty87}. Far down the tip one can
clearly observe competition between branches which gives rise to a
coarsening effect.
In order to study the sidebranching induced by noise we have
measured the half--width $h_{z}(t)$ of the dendrite at various
distances $z$ behind the tip as a function of time. The
half--width of the dendrite as a function of time and its power
spectrum $P_{z}(f)$ are shown in Fig. \ref{fig42b} for a distance
$z=100$ grid points. The data used to compute the power spectrums
were six times the shown lengths. We also computed the
cross--correlation function
\begin{equation}
C(t')=<[h_{Lz}(t+t')-\bar{h}_{Lz}][h_{Rz}(t)-\bar{h}_{Rz}]>
/\sigma_{L} \sigma_{R},
\end{equation}
where $h_{Lz}(t)$, $h_{Rz}(t)$, $\sigma_{L}$ and $\sigma_{R}$ are
the half--width functions and their standard deviations for the
two sides of the dendrite at the same distance from the tip. We
found that $C(0)$ is around $0.4$ for points very close to the tip
and that its value decreases very quickly to $0$ when increasing
$z$. The same behavior was observed in the experiments in Ref.
\cite{dougherty87}.

\begin{figure}
\begin{center}
\includegraphics[width=6cm]{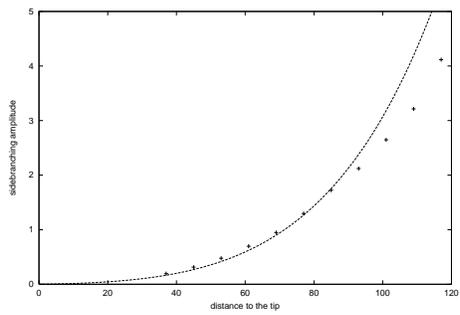}
\caption{\small Square root of the area under the spectral peak as
a function of $z$. \label{fig42c} }
\end{center}
\end{figure}

In Fig. \ref{fig42c} we show the logarithm of the square root of
the area under the spectral peak as a function of $z$. This
representation gives the behavior of the amplitude of the
sidebranching as a function of the distance to the tip. The
amplitude has a different behavior for $z$ up to $100$ (linear
regime) and for values greater than this one (nonlinear regime).
In the region near the tip the amplitude increases exponentially
with $z^{2/5}$, while far down from the tip the growth rate of the
amplitude is slower. Thus, the behavior of the data obtained in
the simulations is consistent with the linear analysis carried out
in Refs. \cite{karma:99,brener95} up to a certain value of $z$.

In conclusion, despite thermal noise could not always be the main
origin of sidebranching, its qualitative characteristics are
common for noise-induced sidebranching independently of its
origin.

\subsection{Mesophase transitions in liquid crystals}

The growth of liquid crystal interfaces in transitions between
mesophases is in many aspects analogous to that of solidification
interfaces. The basic description is expected to be the same,
lying the main differences in the parameter ranges, which often
makes the liquid crystal case particularly suitable from an
experimental point of view. From a theoretical point of view, a
significant parameter difference with respect to the
solidification case is the one associated to the diffusion
coefficients, which are of the same order of magnitude in the two
liquid crystal phases. Another distinct feature is the presence of
strong anisotropies, both at the interface, which can be faceted,
and in the transport properties in the bulk. In this Section we
will show how these anisotropies, introduced into a PFM, can
account for the morphologies obtained in different growth
conditions.

\subsubsection{Anisotropic surface tension}

We have considered a liquid crystalline substance, CCH3, which
presents a N--SmB phase transition at temperature $T_{NS}$. The
pattern formation during the N-SmB phase transition can be
observed experimentally in a quasi two--dimensional geometry
\cite{gonzalezCinca:96,katona:96}. The sample was initially set
above the phase transition temperature $T_{NS}$, and was suddenly
undercooled below this temperature in such a way that small
smectic--B germs nucleated and started to grow.

From the equilibrium shape one can obtain the corresponding
angular dependence of the normalized surface tension $\eta (\theta
)={\sigma(\theta) \over \sigma(0)}$ (being $\theta =0$ the
direction parallel to the smectic layers) by means of the
Wulff--construction \cite{wulff:01}. The function for CCH3 is
\cite{gonzalezCinca:96,katona:96}:
$\eta (\theta) = 1.00 - 0.35 \theta ^2 + 0.01
\theta ^4$ 
in the range $|\theta|\le \pi/2$, while $\eta(\theta) =
\eta(\pi-\theta)$. Clearly $\eta(\theta)$ has cusps at $\theta =
\pm \pi/2$.

Out of equilibrium, CCH3 presents three different morphologies,
ranging from a slightly deformation of the equilibrium shape, to a
butterfly-like morphology and, finally in the largest undercooling
regime, to a four-fold dendritic shape
\cite{gonzalezCinca:96,katona:96,rgc:th}.

The nematic -- smectic--B (N--SmB) phase transition is of first
order, therefore standard models for solidification can be
applied. Like in the previous section, we have employed the
anisotropic PFM of Ref. \cite{wang:93}. In simulations the kinetic
term was taken isotropic. The employed numerical procedure was
described in \cite{gonzalezCinca:96}. We have simulated one
quarter of the full experimental system by locating the initial
smectic--B seed in the lower left corner. Symmetrical (reflecting)
boundary conditions for $\phi$ and $u$ have been imposed on the
four sides of the system.

\begin{figure}
\begin{center}
\includegraphics[width=6cm]{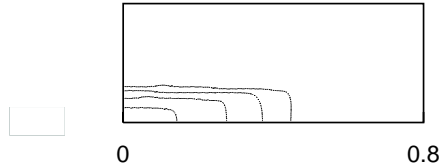}
\caption{\small    \label{fig43a} CCH3. N--SmB interface (points
with $\phi=0.5$) obtained from simulation at subsequent times.
$\Delta = 0.05$.}
\end{center}
\end{figure}

\begin{figure}
\begin{center}
\includegraphics[width=6cm]{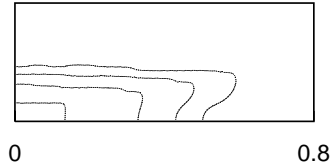}
\caption{\small \label{fig43b} CCH3. N--SmB interfaces
($\phi=0.5$) obtained from simulation at subsequent times. $\Delta
= 0.09$.}
\end{center}
\end{figure}

\begin{figure}
\begin{center}
\includegraphics[width=6cm]{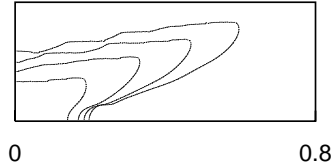}
\caption{\small      \label{fig43c} CCH3. N--SmB interfaces
($\phi=0.5$) obtained from simulation at subsequent times. $\Delta
= 0.2$.}
\end{center}
\end{figure}

\begin{figure}
\begin{center}
\includegraphics[width=6cm]{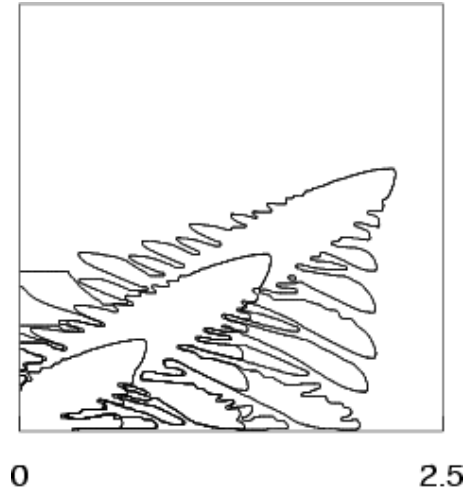}
\caption{\small   \label{fig43d} CCH3. N--SmB interface
($\phi=0.5$) obtained from simulation at subsequent times. $\Delta
= 0.7$.}
\end{center}
\end{figure}

The four main morphologies of CCH3 have been computationally
reproduced on a qualitative level. At small dimensionless
undercooling, simulations show a very slow growth of the germ,
which at the observed times maintains a rectangle--like shape very
similar to the equilibrium one in experiments, {\it i.e.}, with
two parallel facets and two rough convex sides (Fig.
\ref{fig43a}). The growth velocity of the facets is smaller than
that of the convex parts of the interface.
At slightly larger undercoolings, the short sides undergo a first
instability from convex to concave (Fig. \ref{fig43b}).
At larger values of the undercooling the faceted sides start to
bend adopting a slightly concave curvature and macroscopic facets
disappear with the four corners opened up, forming a
butterfly--like shape (Fig. \ref{fig43c}).
Finally, a completely developed four-fold dendrite is obtained in
the large undercooling regime (Fig. \ref{fig43d}).

\subsubsection{Anisotropic heat diffusion}

It has been proved in the previous Section the important
role played by the surface tension anisotropy in the selection of
the steady--state shape of a growing interface. Aside from this
one and kinetic anisotropy, systems such as liquid crystals do
present considerable anisotropies in their transport coefficients
\cite{rondelez:78,kruger:82,zammit:90,melo:90} and their effects
are potentially significant in interfacial growth phenomena. We
address here the growth of dendrites subject to anisotropy in the
heat diffusion coefficient, and its consequences on the growth
morphologies \cite{gonzalezCinca:98a}.

When the anisotropic thermal diffusion is implemented in a PFM,
the $\phi$--equation does not change and the $u$--equation should
be slightly modified to consider a diffusion matrix $D_{ij}$. The
numerical details can be found in Refs.
\cite{gonzalezCinca:98a,rgc:th}.

The experimental system used to study the thermal diffusion
anisotropy is the liquid crystalline substance CCH4, and in
particular its N--SmB phase transition. At large undercoolings,
some SmB seeds whose director directions do not match that of the
sorrounding N phase, present a non-reflection symmetry in their
growth.

\begin{figure}
\begin{center}
\includegraphics[width=6cm]{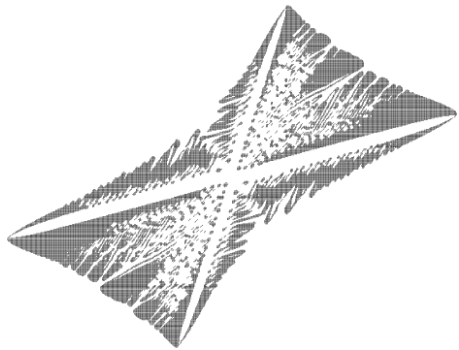}
\caption{\small  \label{fig43e} Simulation for the non--reflection
symmetry.}
\end{center}
\end{figure}

 With the modified PFM, it is possible to reproduce such
experimental morphologies. This is observed in the numerical
simulation (Fig. \ref{fig43e}). Although the used set of
parameters could differ from those real material ones, the
qualitative resemblance with experimental results is remarkably
good \cite{gonzalezCinca:98a}. Axis $x,y$ has been placed along
the principal directions of the diffusion matrix, in order to have
only diagonal terms in the equation. Anisotropy in the diffusion,
defined as $(D_{max}-D_{min})/D_{min}$, has taken a value of
$0.5$, very similar to experimental estimates. The surface free
energy has been rotated in the $(x,y)$ plane, accounting for
different orientations of the director in each phase.
It can be derived from Fig. \ref{fig43e} that the reflection
symmetry has been broken by the inclusion of the anisotropic heat
diffusion (previous simulations with rotated surface tension
function and isotropic heat diffusion did not show the asymmetry
in the growth velocities). In both experiments and simulations the
most developed branches are systematically those growing in the
direction of lowest heat diffusion. Therefore one can conclude
that heat diffusion anisotropy favours dendritic growth in the
lowest diffusion directions. This means that the relevant heat
diffusion process is the one that occurs in the direction
perpendicular to the axis of the dendrite.

\subsection{Results on Viscous Fingering}

The purpose of the present Section is to illustrate the practical use of the
PFM in the case of viscous fingering, with direct quantitative
tests of the thin-interface approximation to assess the usefulness of the
approach in this case. Indeed, the fact that the model has the correct
sharp-interface limit does not guarantee its practical usefulness for several
reasons. On the one hand, the stability of both the bulk phases and the kink
profile must be assured, since this might not be the case in general. On the
other hand, a direct empirical test is necessary to determine quantitatively
how close a finite $\epsilon$ situation is to the sharp-interface limit. This
means finding a set of explicit quantitative criteria to choose all the
nonphysical parameters in order to ensure a desired accuracy. Finally, it is
interesting to find to what extent that model can provide quantitative results
with reasonable computing efforts in actual simulations.

\subsubsection{Linear dispersion relation}
\label{secrd} The first situation in which the model can be tested
is the linear regime of a perturbed planar interface. The linear
dispersion relation has been computed for vanishing viscosity
contrast ($c=0$). The sharp-interface model predicts a linear
growth that does not depend on $c$. However, the phase field model
should exhibit some dependence in the viscosity contrast related
to the finite-$\epsilon$ and -$\tilde{\epsilon}$ corrections (see
Sec. IV in Ref. \cite{vf1}).

We use a single mode occupying the whole channel width (i.e., of
wavelength 1 and wavevector $k=2\pi$) and then vary
the dimensonless surface tension $B$ in order
to change the growth rate of that mode according to the Hele--Shaw
dispersion relation
\begin{equation}
\label{eq:rdhs} \omega_0=|k|(1-Bk^2).
\end{equation}
This is physically completely equivalent to fixing the surface tension
and varying the wavevector of the mode, as can be seen through the
rescaling $k'=\sqrt B k, \omega'_0=\sqrt B \omega_0=|k'|(1-k'^2)$. However,
it is numerically more convenient, since it allows one to use the same
value of $\epsilon$ for all the modes, because $k$ is fixed.
Thus, according to the finite-$\epsilon$
and -$\tilde{\epsilon}$ dispersion relation
derived in Sec. IV of Ref. \cite{vf1},
\begin{equation}
\label{eq:rdthin}
\omega=\omega_0(\frac{1}{\sqrt{1+\frac{\tilde{\epsilon}\omega}{k^2}}}
-\epsilon|k|{\sqrt 2}\frac{5}{6})+{\cal O}(c^2)+{\cal
O}(\epsilon^2).
\end{equation}
We have used $\epsilon=0.01$, $\tilde{\epsilon}=0.1$ for the
deviation from the sharp-interface one, $\omega_0$, to keep below
a 10\%.
\begin{figure}
\begin{center}
\includegraphics[width=7cm]{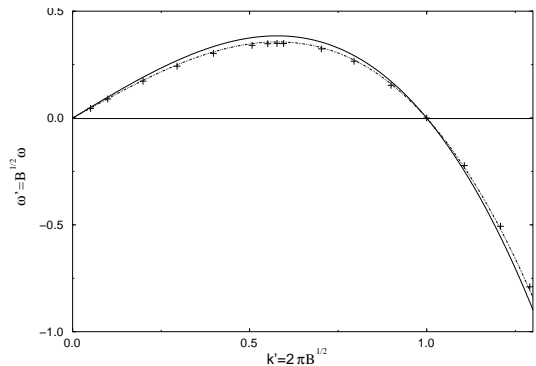}
\caption{\small Linear dispersion relation in scaled  variables
\label{roger-f2} }
\end{center}
\end{figure}

\begin{figure}
\begin{center}
\includegraphics[width=7cm]{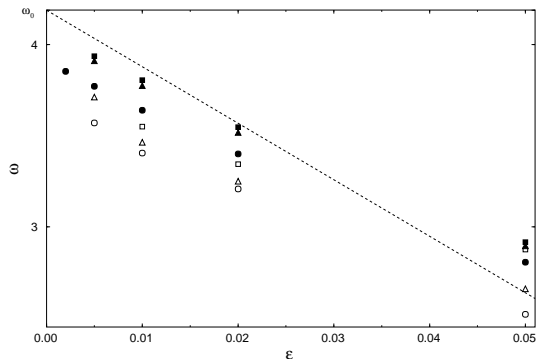}
\caption{\small Convergence to the Hele-Shaw value for the maximum of the
linear dispersion relation curve 
\label{roger-f3} }
\end{center}
\end{figure}

In Fig. \ref{roger-f2} we present the linear dispersion relation thus obtained
in scaled variables. The points (+) correspond
to the measured growth rates for roughly a decade in the
amplitude. Their deviation from the sharp-interface result of Eq.
(\ref{eq:rdhs}) (solid line) keeps below the desired 10\% error
and is fairly well { quantitatively} predicted by the
thin-interface dispersion relation of Eq. (\ref{eq:rdthin})
(dotted line). This quantitative agreement between theory and
numerics is quite remarkable if we take into account that the
thin-interface model is based on an asymptotic expansion in
$\epsilon$. This good agreement is indeed an indication that the
value of $\epsilon$ used is in the asymptotic regime of the
sharp-interface limit, as we will see more clearly in Fig. \ref{roger-f3}.

The growth rate values shown in Fig. \ref{roger-f2} could still be refined by
further decreasing $\epsilon$ and $\tilde{\epsilon}$. This is not
only a theoretical possibility, but can also be done in practice
as we show in Fig. \ref{roger-f3}, although the computation time increases as
explained above. Here, we study the convergence of the growth rate
$\omega$ (y-axis) for the maximum of $\omega'(k')$
($B=8.443...\times 10^{-3}$) to the Hele--Shaw result
$\omega_0=4.188...$ (left upper corner) as we decrease $\epsilon$
(x-axis) and $\tilde{\epsilon}$ (various symbols). The empty
symbols have been obtained with $\Delta x=\epsilon$, whereas the
filled ones correspond to $\Delta x=\epsilon/2$. The growth rates
obtained with $\Delta x=\epsilon$ are always below the ones for
$\Delta x=\epsilon/2$, probably because of the stabilizing effect
of the mesh size. If $\Delta x$ is further decreased the
differences with the values computed with $\Delta x=\epsilon/2$
are tiny, whereas the gap between the $ \Delta x=\epsilon$ and the
$\Delta x=\epsilon/2$ points is pretty large (clearly more than
the differences between distinct symbols ---distinct
$\tilde{\epsilon}$ values--- or adjacent values of $\epsilon$).
This means that the discretization has practically converged to
the continuum model for $\Delta x=\epsilon/2$, but not for $\Delta
x=\epsilon$. That is the reason why we have used $ \Delta
x=\epsilon/2$ in Fig. \ref{roger-f2}.

Moreover, the $\Delta x=\epsilon/2$ points should be described by the
thin-interface model of Eq. (\ref{eq:rdthin}), and this is indeed the case for
small enough values of $\epsilon$. To visualize this we have plotted the
thin-interface prediction for $\tilde{\epsilon}=0$ (dashed line), which is, of
course, a straight line in $\epsilon$. Each set of points with a same
$\tilde{\epsilon}$ value clearly tends to align parallel to this line as
$\epsilon$ decreases ---as Eq. (\ref{eq:rdthin}) predicts---, whereas they
curve up and even cross the line for large values of $\epsilon$, for which we
are beyond the asymptotic regime of validity of Eq. (\ref{eq:rdthin}),
apparently ending near $\epsilon=0.01$. This makes $\epsilon=0.01$ very
suitable for simulations, and confirms it to be within the asymptotic regime as
we pointed out above. Note as well how the growth rate increases with
decreasing values of $\tilde{\epsilon}$ within the same value of $\epsilon$
(vertical columns of points), and how it approaches the dashed line in good
agreement with the values predicted by Eq. (\ref{eq:rdthin}) for values of
$\epsilon$ within the asymptotic regime.
\begin{figure}
\begin{center}
\includegraphics[width=7cm]{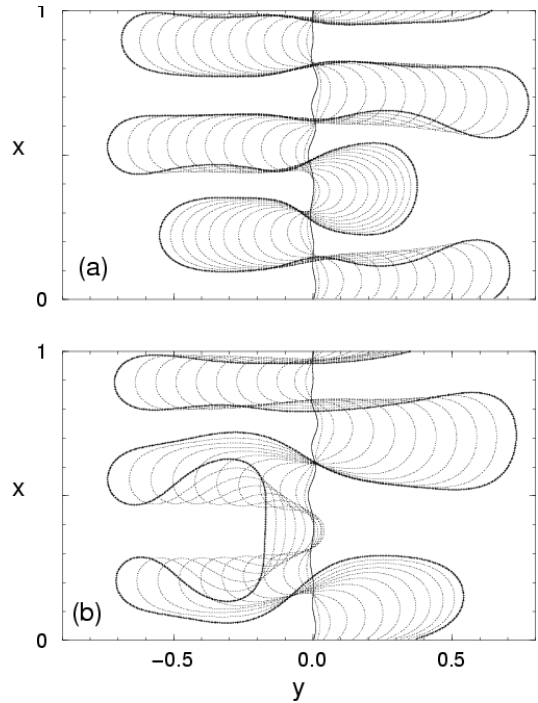}
\caption{\small Time evolution of the interface a) $c=0$
b) $c=.8$
\label{roger-f5} }
\end{center}
\end{figure}

\subsubsection{Finger dynamics}

\label{secst}

The quantitative test in the nonlinear regime are not so simple due to the lack
of exact analytical solutions. For the case of asymptotic Saffman--Taylor
fingers, for instance, only partial numerical information is available, 
in particular for varying viscosity contrast. A detailed comparison
with existing evidence was carried out in Ref. \cite{vf2}

At a qualitative level, however, it is illustrative to plot an example of
multifinger dynamics, in particular to emphasize the importance to have the
viscosity contrast as an explicit parameter in the model. Indeed, the scenario
of finger competition in the high viscosity contrast limit, where larger
fingers screen out the smaller ones, is actually valid only for $c$ very close
to $1$. In contrast, the persistence of the small fingers, which is
characteristic of low viscosity contrasts, seems to be the generic case
\cite{paunethesis}.

Here we use a somewhat experimentally realistic initial
condition consisting of a superposition of sinusoidal modes
with random, uniformly distributed
amplitudes between -0.005 and +0.005 for each wavelength
$\lambda=1,\frac{1}{2},\frac{1}{3}...\frac{1}{7}$ ---i.e., in the
linear regime and random phases. For $B=10^{-3}$ the most amplified of these
wavelengths will be $\lambda=\frac{1}{3}$, so that we expect
3 unequal fingers to appear and there is a chance for mode
interaction and competition
to set in. Wavelengths below
$\lambda=0.161...$ are stable and will decay. We include some of them anyway.
Then all modes are added up to
find the interface position. The stream function predicted by the
linear theory is also obtained by adding up the stream function of each mode,
but all with their peaks centered at the same
final interface position, to avoid the formation of more than one peak of
the stream function across the interface.

Since harmonics of the channel width are present, we have to
refine the $\epsilon$ used in the linear regime. We use
$\epsilon=0.00625$, with
$\Delta x=\epsilon$ to save computation time. The value of
$\tilde{\epsilon}$ is quite crude ($\tilde{\epsilon}=0.5$)
for the equal viscosities case $c=0$, Fig. \ref{roger-f5}a, and especially for
the high viscosity contrast ($c=0.8$) run Fig. \ref{roger-f5}b
($\tilde{\epsilon}=0.2$).

The results are shown in Figs. \ref{roger-f5}a,b for $0.15<t<1.25$ at 
constant
time intervals 0.1 (dots). The last interface is emphasized in
bigger points (+) and the solid line corresponds to the initial
condition for $t=0$. As we can see, the initial condition happens
to have six maxima. Rather quickly, only three of them are left out as
predicted by linear stability, even before entering the non-linear
regime, in which these maxima elongate into well developed
fingers. For vanishing viscosity contrast ($c=0$, Fig. \ref{roger-f5}a), 
there
is no apparent competition, in agreement with experimental
\cite{maher} and numerical
 \cite{aref1,jaume,paunethesis} evidence. Longer and shorter
fingers all advance. Shorter fingers might not advance so quickly,
but they expand to the sides, so they clearly keep growing.
In contrast, starting with the same initial condition that for
$c=0$, the $c=0.8$ run (Fig. \ref{roger-f5}b) shows competition between 
fingers
of the less viscous fluid advancing into the more viscous one,
as it is known to happen in
the Saffman--Taylor problem. The shorter finger now also expands
laterally, but it soon begins to move backwards as a whole. Longer
times may lead to the pinch-off of droplets in both cases 
\cite{paunethesis}.

In summary the basic criteria to control the closeness to the sharp interface
limit are $\epsilon\kappa<<1$,
$\frac{\tilde{\epsilon}\omega}{(1\pm c)k^2}<<1$. More precisely,
we find numerically that the thin-interface model is accurate in
the linear regime with an error below below 10\% if one satisfies
the conditions $\epsilon k \leq 0.06$,
$\frac{\tilde{\epsilon\omega}}{(1\pm c)k^2}\leq 0.016$. The method
could be made more efficient by using an adaptive mesh or
(possibly) by cancelling out the corrections to the
sharp-interface equations remaining (i.e., other than the
Allen--Cahn law)
 in the thin-interface model. For high viscosity contrasts,
$c\sim 1$, a distinct model could possibly be more efficient. From the recent
progress in boundary-integral methods \cite{shelleyreview}, however, the use of
PFM may not be significantly advantageous in this problem except
maybe for the natural incorporation of interface pinch-off. Nevertheless, the
application of the PFM approach in three-dimensional viscous flows
in porous
media, is yet an unexplored and promising future direction.

\section{Conclusions}

Phase-field modeling of interface dynamics in nonequilibrium systems has proved
to be a very useful tool.
In this chapter we have reviewed different aspects of this
active field of research. Phase-field models introduce an auxiliary field to
allow for a diffuse interface. This additional field is coupled to the physical
fields in such a way that the macroscopic, sharp-interface boundary conditions
are effectively reproduced. In this way, a free-boundary problem is transformed
into a set of partial differential equations, usually much simpler to handle.
In this chapter a phase-field model for solidification has been explicitly
derived from a free energy. As a second example, a phase-field model for the
Saffman--Taylor problem has been presented. In this case the derivation is not
based on a free-energy functional. The thin-interface approximation, which
includes corrections on the interface thickness, have been discussed for these
two examples. We have also presented some illustrative applications such as the
study of fluctuations in directional solidification of alloys and dendritic
sidebranching in free solidification, growth of mesophases in liquid crystals
and the dynamics of viscous fingering.
 
At the present level of development of phase-field model techniques, these have
been shown to be quantitatively accurate with reasonable computational cost,
and usually advantageous with respect to other techniques in different aspects.
Phase-field models are simple to be implemented in a computer and contain a
complete description of the full nonlinear and nonlocal properties of the
macroscopic free-boundary problems. In addition, in the phase-field framework
it is relatively simple to introduce additional perturbations, such as
fluctuations and static disorder, or modifications of the boundary conditions.
Future lines of research will certainly include quantitative studies of
interfaces in three dimensions, the consideration of complex fluids and
biological systems, and possibly applications at the nanoscale. Undoubtly the
phase-field approach holds the promise of fruitful applications to new problems
in years to come.


\end{document}